\ifdefined\TCOMreviewversion
\documentclass[letterpaper,12pt,onecolumn,draftclsnofoot]{IEEEtran}
\else
\documentclass[letterpaper,journal]{IEEEtran}
\fi

\usepackage[T1]{fontenc}
\usepackage{amsmath,amssymb,amsfonts,mathtools}
\usepackage{graphicx}
\usepackage{pgfplots}
\usepackage{pgfplotstable}
\pgfplotsset{compat=1.18}
\usepgfplotslibrary{statistics,groupplots,fillbetween}
\usepackage{booktabs}
\usepackage{multirow}
\usepackage{array}
\newcolumntype{L}[1]{>{\raggedright\arraybackslash}p{#1}}
\usepackage{cite}
\usepackage{xcolor}
\usepackage{microtype}
\usepackage{xurl}
\ifCLASSOPTIONtwocolumn\usepackage{flushend}\fi
\usepackage{subcaption} 

\newtheorem{theorem}{Theorem}
\newtheorem{lemma}[theorem]{Lemma}
\newtheorem{corollary}[theorem]{Corollary}
\newtheorem{proposition}[theorem]{Proposition}
\newtheorem{definition}[theorem]{Definition}
\newtheorem{construction}[theorem]{Construction}
\newtheorem{remark}[theorem]{Remark}
\newtheorem{example}[theorem]{Example}
\newenvironment{proof}{\begin{IEEEproof}}{\end{IEEEproof}}

\newcommand{\PI}{P_{\mathrm{I}}}
\newcommand{\PS}{P_{\mathrm{S}}}
\newcommand{\Csub}{\mathcal{C}_{\mathrm{sub}}}
\newcommand{\Gsub}{\mathbf{G}_{\mathrm{sub}}}
\newcommand{\wt}{\mathrm{wt}}
\newcommand{\sv}{\mathbf{s}}
\newcommand{\tv}{\mathbf{t}}
\newcommand{\kv}{\mathbf{k}}
\newcommand{\cv}{\mathbf{c}}
\newcommand{\uv}{\mathbf{u}}
\newcommand{\Ftwo}{\mathbb{F}_2}
\newcommand{\PSsub}{P_{S,\mathrm{sub}}}
\newcommand{\PSord}{P_{S,\mathrm{ord}}}
\newcommand{\PSind}{P_{S,\mathrm{ind}}}

\begin{document}

\title{Code-Based Message Authentication with Statistical Guarantees via Coordinate Selection and Masking}

\author{Hsuan-Po~Liu,~\IEEEmembership{Student Member,~IEEE}
and~Hessam~Mahdavifar,~\IEEEmembership{Member,~IEEE}
\thanks{This paper was presented in part at the IEEE International Symposium on Information Theory (ISIT), Athens, Greece, July 2024 \cite{liu2024projective}.
This work was supported by the Center for Ubiquitous Connectivity (CUbiC), sponsored by Semiconductor Research Corporation and the Defense Advanced Research Projects Agency under the JUMP 2.0 program.}
\thanks{H.-P.~Liu is with the Department of Electrical Engineering and Computer Science, University of Michigan, Ann Arbor, MI 48109 USA (e-mail: hsuanpo@umich.edu).
H.~Mahdavifar is with the Department of Electrical and Computer Engineering, Northeastern University, Boston, MA 02115 USA, and also with the Department of Electrical Engineering and Computer Science, University of Michigan, Ann Arbor, MI 48109 USA (e-mail: h.mahdavifar@northeastern.edu).}}

\maketitle

\begin{abstract}
We study message authentication over a public channel against a computationally
unbounded adversary. Motivated by low-complexity applications, we consider schemes that leverage
off-the-shelf binary linear codes already used for error correction in communication systems. Existing code-based constructions derive an authentication tag primarily from a single symbol in a $q$-ary code, which limits the tag length. However, selecting several binary coordinates allows a fixed binary linear code to support different tag lengths. We characterize how coordinate selection and ordering affect the statistical guarantees against impersonation and substitution attacks. In particular, we explore three methods: the subset selector, the ordered selector, and the independent selector.
They read uniformly chosen coordinates in a fixed public or a secret random order, or with replacement, and add a uniform mask.
With a fresh key per message, all three achieve the optimal impersonation probability $2^{-l_t}$ for $l_t$-bit tags.
We bound the subset selector's substitution probability using only nonzero weights.
Those of the other two depend only on the code length $n$, $l_t$, and the nonzero weight farthest from $n/2$.
The independent selector achieves $2^{-l_t}$ for substitution when all nonzero weights equal $n/2$.
Finally, we evaluate first-order Reed--Muller and Bose--Chaudhuri--Hocquenghem dual subcodes.
\end{abstract}

\ifCLASSOPTIONonecolumn\noindent\begin{minipage}{\textwidth}\fi
\ifCLASSOPTIONonecolumn\end{minipage}\par\fi

\section{Introduction}

\IEEEPARstart{A}{uthentication} enables a receiver to verify both the origin and integrity of a message sent over a public channel.
It is a basic requirement in communication systems \cite{hakeem2022security}.
Authentication codes, introduced by Gilbert, MacWilliams, and Sloane \cite{gilbert1974codes} and formalized by Simmons \cite{simmons1985authentication}, provide statistical guarantees against computationally unbounded adversaries.
As such, they complement post-quantum cryptography, whose security relies on computational assumptions \cite{campagna2015quantum}.
Inspired by the growing role of error correction codes in post-quantum security systems, such as Hamming Quasi-Cyclic (HQC) \cite{alagic2025status}, we study systematic authentication using off-the-shelf binary linear error correction codes from a fresh perspective.
From the implementation perspective, a major motivation behind this approach is to reuse existing modules for physical layer communication \cite{liu2026locally}. More specifically,
since channel encoders are already implemented in communication systems, an authentication code that computes its tags from codewords can potentially reuse these encoders.

In a systematic authentication code (Fig.~\ref{fig:sys}), assuming a shared secret key, the sender transmits the source in clear appended with an authentication tag computed from the source.
We focus on statistical guarantees against two standard attacks, namely, \emph{impersonation} and \emph{substitution} \cite{simmons1985authentication}.
In impersonation, the adversary forges a source--tag pair without observing a transmission.
In a substitution attack, the adversary intercepts a valid source–tag pair and uses what it observes to choose a tag for a different source. It sends this replacement pair to the receiver, and the attack succeeds if the receiver accepts the new source as authentic.
As in the classical model, each key is used to authenticate one source.
Let $\PI$ and $\PS$ denote the largest probabilities that a forged pair is accepted in impersonation and substitution, respectively.
For an $l_t$-bit tag, both satisfy a statistical information-theoretic lower bound $\PI,\PS\geq 2^{-l_t}$.

Authentication codes have been constructed from combinatorial designs
\cite{stinson1992combinatorial}, geometric codes
\cite{bierbrauer1994families,bierbrauer1997universal}, highly nonlinear
functions \cite{chanson2003cartesian,ding2004systematic,carlet2006authentication},
exponential sums \cite{helleseth1996universal}, algebraic curves
\cite{xing2000constructions}, Galois rings \cite{ozbudak2006some}, and
error-correcting codes \cite{johansson1994relation,kabatianskii1996cardinality,wang2003linear,ding2005coding,ding2007generic,liu2018new}.
Authentication codes can also be obtained from universal hash families
\cite{wegman1981new,stinson1994universal}, which are typically designed for
short keys. In the earlier code-based constructions summarized in
Table~\ref{tab:priorwork}, the key selects a single coordinate of a codeword over
$\mathbb{F}_q$, and the corresponding symbol, optionally masked, forms the
tag. Hence, the tag has $q$ possible values, and $\PS\geq1/q$.
In particular, a binary code yields only a one-bit tag, and an $l_t$-bit tag
requires a code over $\mathbb{F}_{2^{l_t}}$, so that changing the tag length
requires changing the code.
Prior code-based constructions
\cite{johansson1994relation,kabatianskii1996cardinality,wang2003linear,ding2005coding,ding2007generic,liu2018new}
do not consider multi-coordinate tags as in this paper, i.e., tags formed from
several coordinates of the same codeword.
Extending our conference paper \cite{liu2024projective}, we form an $l_t$-bit
tag from $l_t$ coordinates of one binary codeword, so that a fixed binary
linear code supports different tag lengths.
The question is then how the choice and order of these coordinates affect the
statistical guarantees of the resulting authentication code.

\begin{table*}[t]
\centering
\caption{Comparison of code-based systematic authentication constructions
in terms of what determines $\PS$ and whether $\PS$ achieves the
tag-space lower bound. For tag space $\mathcal T$, every systematic
authentication code satisfies $\PS \geq \PI \geq 1/|\mathcal T|$.
The masked constructions achieve $\PI = 1/|\mathcal T|$.
The row for \cite{liu2018new} uses \cite[Thm.~2]{liu2018new} with the
exponent parameter set to one. Here $n$ is the code length, $l_s$ is the
dimension of the binary subcode, and $\rho$ is its maximum weight
imbalance, defined in \eqref{eq:rho} of Section~\ref{sec:ordered}.
}
\label{tab:priorwork}
\ifCLASSOPTIONtwocolumn
\scriptsize
\setlength{\tabcolsep}{3pt}
\else
\scriptsize
\setlength{\tabcolsep}{3pt}
\fi
\renewcommand{\arraystretch}{1.3}
\begin{tabular}{@{}L{0.165\textwidth}L{0.095\textwidth}L{0.175\textwidth}L{0.17\textwidth}L{0.115\textwidth}L{0.21\textwidth}@{}}
\toprule
Construction & Tag & Secret key & $\PS$ determined by & Lower bound on $\PS$ & Lower bound achieved\\
\midrule
Projective \cite[Thm.~1]{ding2005coding}
 & one $\mathbb F_q$ symbol & one coordinate
 & co-occurrences of symbol pairs in two codewords
 & $1/q$, fixed by the code alphabet
 & with the generalized first-order RM subcode \cite[Thm.~12]{ding2005coding}\\
Generic Cartesian \cite[Thm.~1]{ding2007generic}
 & one $\mathbb F_q$ symbol & one coordinate and a mask in $\mathbb F_q$
 & largest symbol frequency in a nonzero codeword
 & $1/q$, fixed by the code alphabet
 & iff every nonzero codeword contains each symbol $n/q$ times\\
\hspace{1em}with three-weight cyclic codes \cite[Thm.~2]{liu2018new}
 & one $\mathbb F_q$ symbol & one coordinate and a mask in $\mathbb F_q$
 & as in \cite{ding2007generic}, counted exactly
 & $1/q$
 & no, $\PS\to1/q$ as the code length grows\\
\midrule
\textbf{This paper}, subset (Constr.~\ref{con:main})
 & $l_t$ bits & $l_t$ distinct coordinates in fixed order and an $l_t$-bit mask
 & projection patterns of nonzero codewords under the fixed order
 & $2^{-l_t}$, set by $l_t$ for a fixed code
 & only if $l_t=1$ and $\rho=0$\\
\textbf{This paper}, ordered (Constr.~\ref{con:ordered})
 & $l_t$ bits & $l_t$ distinct coordinates in secret order and an $l_t$-bit mask
 & $n$, $l_t$, and $\rho$ only
 & $2^{-l_t}$
 & only if $l_t=1$ and $\rho=0$, approached as $n\to\infty$ if $\rho=0$\\
\textbf{This paper}, independent (Constr.~\ref{con:iid})
 & $l_t$ bits & $l_t$ coordinates sampled with replacement and an $l_t$-bit mask
 & $l_t$ and $\rho$: $((1+\rho)/2)^{l_t}$
 & $2^{-l_t}$
 & iff $\rho=0$, which requires $l_s\leq\log_2 n$, e.g., the first-order RM subcode (Section~\ref{sec:rm})\\
\bottomrule
\end{tabular}
\end{table*}

We address this question by comparing three rules for selecting $l_t$ tag bits
from a binary codeword. The \emph{subset selector} reads a uniformly chosen
subset of $l_t$ coordinates in a fixed public order. The \emph{ordered
selector} chooses the subset in the same way but reads its bits in a secret,
uniformly random order. And the \emph{independent selector} draws $l_t$
coordinates independently and uniformly with replacement and reads the bits
in the order drawn. In each case, the sender masks the selected bits with
an independent uniform pad from the readily available shared secret key to form the authentication tag. For each selector, we study $\PI$ and $\PS$.

\subsection{Related Work}
\label{ssec:prior}

Table~\ref{tab:priorwork} compares three earlier constructions whose
substitution guarantees can be expressed in terms of code structure.
In the projective construction of \cite{ding2005coding}, $\PS$ depends on
how often a given pair of symbols occurs at corresponding coordinates of
two codewords. A generalized first-order Reed--Muller (RM) subcode achieves
$\PI=\PS=1/q$ \cite[Thm.~12]{ding2005coding}. In the masked Cartesian
construction of \cite{ding2007generic}, the mask cancels in the difference
between two tags. Therefore, for the full linear construction, $\PS$ equals
the largest relative frequency of any symbol in a nonzero codeword.
Counting these symbol occurrences gives exact values of $\PS$ for the
three-weight cyclic codes studied in \cite{liu2018new}.

When $l_t=1$, all three of our selectors reduce to the binary instance of
the masked construction in \cite{ding2007generic}. For longer tags, uniform
masking still gives $\PI=2^{-l_t}$, but a successful substitution requires
the adversary to predict an entire $l_t$-bit pattern selected from a
nonzero codeword. The probability of predicting that pattern depends on
how the coordinates are sampled and ordered. For example, permuting the
coordinates preserves the weight of every codeword, yet can change $\PS$
for the subset selector.

\subsection{Our Contributions}

Our conference paper \cite{liu2024projective} analyzed fixed-order subset
selection for RM subcodes. Here we extend the analysis to arbitrary
binary linear subcodes and compare three proposed selection rules. All three achieve
$\PI=2^{-l_t}$ assuming a fresh uniform key per message. Hence, our main results concern $\PS$ and are summarized as follows:

\begin{figure*}[t]
\centering
{\includegraphics[width=.93\textwidth]{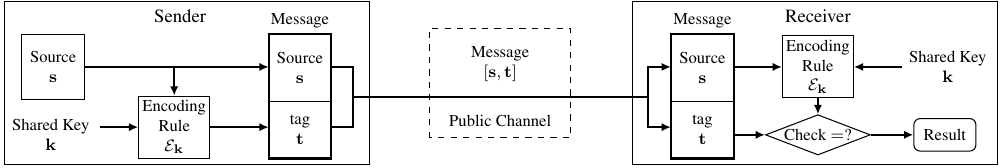}}\\
  \caption{System architecture of the proposed systematic authentication scheme.}\label{fig:sys}
\end{figure*}

\begin{itemize}
\item \emph{Coordinate order.} The subset selector chooses $l_t$ distinct
coordinates uniformly and reads them in a fixed public order. We bound
$\PS$ using only the nonzero codeword weights. The bound holds for every
public ordering of the $n$ coordinates, and at least one ordering achieves
it for each subcode.

\item \emph{Other selectors and comparison.} The ordered selector reads a uniformly
chosen subset in a secret random order, while the independent selector
draws coordinates uniformly with replacement. For fixed $n$ and $l_t$,
we derive exact expressions for $\PS$ in terms of the maximum weight
imbalance, the largest value of $|2w/n-1|$ over nonzero codeword weights
$w$. These expressions identify when each selector gives the smaller $\PS$.

\item \emph{Balance and code choice.} A subcode is exactly balanced if
every nonzero codeword has weight $n/2$. For $2\leq l_t<n$, only the
independent selector can achieve $\PS=2^{-l_t}$, and it does so exactly
under this balance condition. We evaluate all three selectors on a balanced
first-order RM subcode and derive bounds for the ordered and
independent selectors on subcodes derived from binary Bose--Chaudhuri--Hocquenghem (BCH) codes.
\end{itemize}

The remainder of the paper is organized as follows.
In Section~\ref{sec:prelim}, we present the model.
Section~\ref{sec:construction} describes the three constructions, which all achieve $\PI=2^{-l_t}$ (Theorem~\ref{thm:PI}).
In Section~\ref{sec:analysis}, we bound $\PS$ of the subset selector using only the nonzero codeword weights (Corollary~\ref{cor:cert}).
We also show that some coordinate order achieves this weight-set bound (Corollary~\ref{thm:minimax}).
Then, in Section~\ref{sec:ordered}, we derive the exact $\PS$ of the other two selectors in terms of the maximum weight imbalance (Theorems~\ref{thm:ordered} and~\ref{thm:iid} and Proposition~\ref{prop:ordered-imbalance}).
Also, the ordered selector's $\PS$ is at most that of the subset selector (Theorem~\ref{thm:sandwich}).
Furthermore, for $2\leq l_t<n$, only the independent selector can achieve $2^{-l_t}$ (Remark~\ref{rem:no-optimum}).
It does so if and only if the subcode is exactly balanced (Corollary~\ref{cor:optimal}).
For $2\leq l_t\leq\lfloor n/2\rfloor$, the ordered selector has a smaller $\PS$ than the independent selector if and only if the subcode is unbalanced (Corollary~\ref{cor:ord-ind-comparison}).
In Section~\ref{sec:rm}, we evaluate the first-order RM subcode exactly (Corollary~\ref{thm:rm}).
Also, exact balance limits the source dimension to $\log_2 n$ (Proposition~\ref{prop:balanced}).
To support larger source dimensions, we bound $\PS$ of the ordered and independent selectors for BCH-dual subcodes (Corollary~\ref{prop:bch-guarantee}).
Section~\ref{sec:eval} provides exact evaluations.
For instance, the coordinate order alone changes $\PS$ of the subset selector by more than three orders of magnitude (Table~\ref{tab:orderings}).
Finally, Section~\ref{sec:conclusion} concludes the paper.

\section{Preliminaries}
\label{sec:prelim}
\begin{figure}[t]
\centering
\begin{subfigure}[b]{0.48\columnwidth}
\centering
\includegraphics[width=\textwidth]{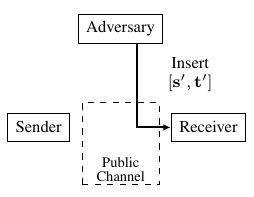}
\caption{Impersonation attack}
\label{fig:PI}
\end{subfigure}
\hfill
\begin{subfigure}[b]{0.48\columnwidth}
\centering
\includegraphics[width=\textwidth]{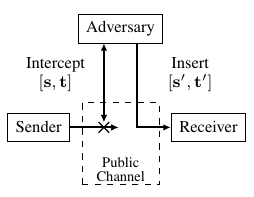}
\caption{Substitution attack}
\label{fig:PS}
\end{subfigure}
\caption{Impersonation and substitution attacks.}
\label{fig:attacks}
\end{figure}

In this section, we describe the systematic authentication model and define the deception probabilities.

\subsection{Systematic Authentication}

A systematic authentication code is a four-tuple $(\mathcal S,\mathcal T,\mathcal K,\allowbreak\{\mathcal E_{\kv}:\kv\in\mathcal K\})$, where $\mathcal S$, $\mathcal T$, and $\mathcal K$ are the source, tag, and key spaces, respectively.
Also, $\mathcal E_{\kv}:\mathcal S\to\mathcal T$ is the deterministic encoding rule for key $\kv$.
Using a shared secret key $\kv$, the sender transmits $[\sv,\tv]$ with $\tv=\mathcal E_{\kv}(\sv)$.
The receiver accepts a received pair if and only if its tag equals the tag recomputed from its source.
The system is summarized in Fig.~\ref{fig:sys}.
We model an adversary that controls the public channel and may mount either of the following two attacks, illustrated in Fig.~\ref{fig:attacks}:
\begin{enumerate}
\item \textbf{Impersonation:} Insert a forged pair $[\sv',\tv']$ without observing a legitimate transmission (Fig.~\ref{fig:PI}).
\item \textbf{Substitution:} Intercept a valid pair $[\sv,\tv]$ and replace it by $[\sv',\tv']$ with $\sv'\neq\sv$ (Fig.~\ref{fig:PS}).
\end{enumerate}
An attack succeeds if the receiver accepts the inserted pair $[\sv',\tv']$, i.e., if this pair is valid with $\tv'=\mathcal E_{\kv}(\sv')$.
Security is measured by the largest success probability of each attack, with unbounded computational power of the adversary.

The adversary knows the authentication code, i.e., the sets $\mathcal S$, $\mathcal T$, and $\mathcal K$ and every encoding rule, but not the key $\kv$.
In our constructions (Section~\ref{sec:construction}), this means that the adversary knows the subcode, its generator matrix, and the distribution of the key.

\subsection{Probabilities of Successful Deception}
\label{ssec:deception}
We assume that the key is uniform on $\mathcal K$ and that $|\mathcal S|\geq2$.
The source is independent of the key, but its distribution need not be uniform, since security is evaluated for every fixed source and valid observed tag.
In impersonation, the adversary observes nothing and may insert any pair $[\sv',\tv']\in\mathcal S\times\mathcal T$, which succeeds if $\tv'=\mathcal E_{\kv}(\sv')$.
Hence, the impersonation probability is defined as
\begin{equation}
\PI:=\max_{\sv',\tv'}\Pr\{[\sv',\tv']\text{ valid}\},
\end{equation}
where the probability over the uniform key is the fraction of keys under which $[\sv',\tv']$ is valid:
\begin{equation}
\Pr\{[\sv',\tv']\text{ valid}\}
=\frac{|\{\kv\in\mathcal K:\mathcal E_{\kv}(\sv')=\tv'\}|}
{|\mathcal K|}.
\label{eq:P_I_prob}
\end{equation}

In substitution, the adversary observes the pair $[\sv,\tv]$ transmitted by the sender, where $\tv=\mathcal E_{\kv}(\sv)$.
The observed pair $[\sv,\tv]$ rules out every key $\kv$ with $\mathcal E_{\kv}(\sv)\neq\tv$.
Then the best strategy is to insert a pair $[\sv',\tv']$ with a different source $\sv'\neq\sv$ that satisfies $\tv'=\mathcal E_{\kv}(\sv')$ with the largest conditional probability given this observation, i.e., under the largest fraction of the remaining keys.
Note that the condition $\sv'\neq\sv$ excludes forwarding the observed pair unchanged, which the receiver always accepts.
Thus, the substitution probability is defined as
\begin{equation}
\PS:=\max_{\sv,\tv}\max_{\sv'\neq\sv,\tv'}
\Pr\{[\sv',\tv']\text{ valid}\mid[\sv,\tv]\text{ observed}\},
\label{eq:P_S_worst}
\end{equation}
where the outer maximum is over all pairs $[\sv,\tv]$ that the sender can transmit, i.e., with $\Pr\{\mathcal E_{\kv}(\sv)=\tv\}>0$, the inner maximum is over all $\sv'\in\mathcal S\setminus\{\sv\}$ and $\tv'\in\mathcal T$, and the conditional probability is
\begin{equation}
\frac{\left|\left\{\kv\in\mathcal K:
\begin{matrix*}[l]
\mathcal E_{\kv}(\sv)=\tv,\\
\mathcal E_{\kv}(\sv')=\tv'
\end{matrix*}\right\}\right|}
{|\{\kv\in\mathcal K:\mathcal E_{\kv}(\sv)=\tv\}|}.
\label{eq:P_S_payoff}
\end{equation}
These worst-case definitions give $\PS\geq\PI\geq|\mathcal T|^{-1}$, as shown in \cite{stinson1992combinatorial}.
Our goal is to minimize both deception probabilities.

\section{The Proposed Constructions}
\label{sec:construction}

In this section, we present three constructions.
Each of them maps sources into a binary linear subcode, selects $l_t$ codeword coordinates, and adds a uniform mask to the selected bits.
They differ only in how the coordinates are selected and in which order they are read.

\subsection{Common Encoding Framework}
\label{ssec:con-framework}

All vectors are row vectors, and all vector and matrix operations are over $\Ftwo$.
Let $\mathcal{C}$ be a binary linear $(n,k)$ code with a full-row-rank generator matrix $\mathbf{G}\in\Ftwo^{k\times n}$, so that $\mathcal{C}=\{\uv\mathbf{G}:\uv\in\Ftwo^k\}$.
For a source dimension $1\leq l_s\leq k$, we select $l_s$ distinct rows of $\mathbf{G}$ to form the matrix $\Gsub\in\Ftwo^{l_s\times n}$.
Note that these rows are linearly independent since $\mathbf{G}$ has full row rank.
Then the codeword of a source $\sv\in\Ftwo^{l_s}$ is
\begin{equation}
\cv_{\sv}=\sv\Gsub.
\label{eq:source-embedding}
\end{equation}
The resulting authentication subcode is then
\begin{equation}
\Csub
=
\{\sv\Gsub:\sv\in\Ftwo^{l_s}\}
\subseteq\mathcal{C}.
\label{eq:subcode}
\end{equation}
Since $\operatorname{rank}(\Gsub)=l_s$, the source-to-codeword map is injective, and $\Csub$ is an $(n,l_s)$ binary linear code containing exactly $2^{l_s}$ codewords.
We require $\mathbf{1}_n\notin\Csub$ and select the rows of $\mathbf{G}$ accordingly, since Proposition~\ref{prop:allones} shows that $\PS=1$ otherwise.
If $l_s=k$, this exclusion requires $\mathbf1_n\notin\mathcal C$.
This structure supports the potential reuse of a compatible encoder that implements $\uv\mapsto\uv\mathbf{G}$.
Then $\sv\Gsub$ is obtained by placing the source bits in the input positions of the selected rows and setting the remaining inputs to zero.
Reusing such an encoder also requires access to the selected output coordinates, and coordinate extraction and masking remain necessary.

Let $[n]:=\{1,\ldots,n\}$, and let the source and tag spaces be $\mathcal{S}=\Ftwo^{l_s}$ and $\mathcal{T}=\Ftwo^{l_t}$, respectively.
We compare the three constructions over the common range $1\leq l_t<n$.
Note that the independent selector and the formula for its substitution probability (Theorem~\ref{thm:iid}) remain valid for every $l_t\geq1$.
All three constructions use the same $\Gsub$ and tag space, and their secret keys are generated independently of the source.
The selected codeword entries are added to a mask $\kv_2\in\mathcal{K}_2=\Ftwo^{l_t}$ that is uniform and independent of the selection.
Note that neither generator matrix is required to be in systematic form.

\subsection{Subset Selector}
The subset selector chooses a uniformly random set of $l_t$ distinct coordinates and reads their bits in increasing order.
More specifically, we represent its selection by a binary vector $\kv_1$ in the set
\begin{equation}
\mathcal{K}_1
=
\{\kv_1\in\Ftwo^n:\operatorname{wt}(\kv_1)=l_t\},
\label{eq:subset-key-space}
\end{equation}
where $\operatorname{wt}(\cdot)$ denotes the Hamming weight.
The selected coordinates form the set
\begin{equation}
\{i\in[n]:(\kv_1)_i=1\}
=
\{i_1,\ldots,i_{l_t}\},
\end{equation}
where $i_1<\cdots<i_{l_t}$.
We then define
\begin{equation}
    \cv_{\sv,\kv_1}:=[c_{\sv,i_1},\ldots,c_{\sv,i_{l_t}}],
\end{equation}
where $c_{\sv,i}$ denotes the $i$th coordinate of $\cv_{\sv}$.
We call $\cv_{\sv,\kv_1}$ the projection of $\cv_{\sv}$ under $\kv_1$, and each of its possible values a projection pattern.

\begin{construction}[Subset selector]
\label{con:main}
Let $\kv_1$ be chosen uniformly from $\mathcal{K}_1$, independently of the uniform mask $\kv_2$.
For the key $\kv=(\kv_1,\kv_2)\in\mathcal{K}=\mathcal{K}_1 \times\mathcal{K}_2$, the encoding rule is
\begin{equation}
\tv
=
\mathcal{E}_{\kv}(\sv)
=
\cv_{\sv,\kv_1}+\kv_2.
\label{eq:enc}
\end{equation}
\end{construction}
Note that although $\kv_1$ is written as an $n$-bit incidence vector, it can be represented compactly by an index among the $\binom{n}{l_t}$ admissible vectors.
Hence, the key length, i.e., the number of bits in a minimum fixed-length representation of the selection and its mask, is
\begin{equation}
L_{\mathrm{sub}}
=
\Bigl\lceil\log_2\tbinom{n}{l_t}\Bigr\rceil+l_t.
\label{eq:keylen}
\end{equation}
Here, the final $l_t$ bits represent the mask.
The direct encoding requires one multiplication by $\Gsub$, the extraction of $l_t$ coordinates, and an $l_t$-bit exclusive-or.

\subsection{Ordered Selector}
\label{ssec:con-ordered}

The ordered selector augments the subset selector with an independent secret permutation.
Let $\Sigma_{l_t}$ denote the set of all permutations of the index set $[l_t]=\{1,\dots,l_t\}$.
Then the key is $\kv=(\kv_1,\sigma,\kv_2)\in \mathcal{K}_1\times\Sigma_{l_t}\times\mathcal{K}_2$, where all three components are independent and uniform.
For the selected coordinates $i_1<\cdots<i_{l_t}$ of $\kv_1$, we define
\begin{equation}
    \sigma(\cv_{\sv,\kv_1}):=[c_{\sv,i_{\sigma(1)}},\ldots,c_{\sv,i_{\sigma(l_t)}}].
\end{equation}

\begin{construction}[Ordered selector]
\label{con:ordered}
The encoding rule of the ordered selector is
\begin{equation}
\tv
=
\mathcal{E}_{\kv}(\sv)
=
\sigma(\cv_{\sv,\kv_1})+\kv_2.
\label{eq:encordered}
\end{equation}
\end{construction}

Note that the pair $(\kv_1,\sigma)$ is in bijection with the ordered $l_t$-tuples of distinct coordinates from $[n]$.
Hence, with the falling factorial $(x)_r:=x(x-1)\cdots(x-r+1)$ and $(x)_0:=1$, the selection of the ordered selector is uniform on a space of cardinality
\begin{equation}
    (n)_{l_t}=\binom{n}{l_t}l_t!.
\end{equation}
The key length of the ordered selector is
\begin{equation}
L_{\mathrm{ord}}
=
\bigl\lceil\log_2 (n)_{l_t}\bigr\rceil+l_t.
\label{eq:keylen-ordered}
\end{equation}
In other words, its selector entropy exceeds that of the subset selector by $\log_2(l_t!)$ bits.
With the ceilings, the difference between the key lengths \eqref{eq:keylen} and \eqref{eq:keylen-ordered} is within one bit of this value.
The encoding operations, the tag length, and the mask length are unchanged.
The only difference is that the coordinates are extracted in the secret order instead of the increasing order.

\subsection{Independent Selector}
\label{ssec:con-independent}

The independent selector chooses each coordinate index independently and uniformly from $[n]$, where repeated indices are allowed.
Then the unmasked tag consists of the corresponding bits in the selected order.

\begin{construction}[Independent selector]
\label{con:iid}
Let $\mathcal{I}=(I_1,\ldots,I_{l_t})$, where $I_1,\ldots,I_{l_t}$ are independent and uniform on $[n]$.
For the key $\kv=(\mathcal{I},\kv_2)$, with $\kv_2$ uniform and independent of $\mathcal{I}$, the encoding rule is
\begin{equation}
\tv
=
\mathcal{E}_{\kv}(\sv)
=
[c_{\sv,I_1},\ldots,c_{\sv,I_{l_t}}]+\kv_2.
\label{eq:enciid}
\end{equation}
\end{construction}

The selection $\mathcal{I}$ is uniform on $[n]^{l_t}$, which contains $n^{l_t}$ elements, and hence its entropy is $l_t\log_2 n$ bits.
The key length of the independent selector is
\begin{equation}
L_{\mathrm{ind}}
=
\bigl\lceil l_t\log_2 n\bigr\rceil+l_t.
\label{eq:keylen-independent}
\end{equation}
Alternatively, storing each index separately uses $\lceil\log_2 n\rceil$ bits per index, i.e., $l_t\lceil\log_2 n\rceil$ bits for the selection, which equals the first term of \eqref{eq:keylen-independent} when $n$ is a power of two.
The key lengths \eqref{eq:keylen}, \eqref{eq:keylen-ordered}, and \eqref{eq:keylen-independent} count the bits needed to store a key of each construction, i.e., its selection and its mask.

\subsection{Common Structural Observations}
\label{ssec:con-observations}

For a fixed selection $\alpha$, let $\varphi_{\alpha}:\Ftwo^n\to\Ftwo^{l_t}$ denote the corresponding coordinate-extraction map, where $\alpha$ is $\kv_1$, $(\kv_1,\sigma)$, or $\mathcal{I}$ for the subset, ordered, or independent selector, respectively.
Note that each map is linear, including when coordinates repeat, and every construction can be written as
\begin{equation}
    \tv=\mathcal{E}_{\kv}(\sv)=\varphi_{\alpha}(\sv\Gsub)+\kv_2.
\end{equation}

\begin{theorem}[Optimal impersonation probability]
\label{thm:PI}
All three constructions achieve $\PI=2^{-l_t}=|\mathcal T|^{-1}$.
\end{theorem}
\begin{proof}
For every fixed source $\sv$, selection $\alpha$, and tag $\tv$, uniform masking gives
\begin{equation}
\Pr_{\kv_2}\{\varphi_{\alpha}(\cv_{\sv})+\kv_2=\tv\}=2^{-l_t}.
\label{eq:masked-tag-uniform}
\end{equation}
Then averaging over the selection gives the same probability for every candidate pair, which meets the lower bound from Section~\ref{sec:prelim}.
\end{proof}
Note that, for each fixed source, \eqref{eq:masked-tag-uniform} shows that the observed tag is independent of the selection.
In other words, observing one valid source--tag pair leaves the selection uniform.

For the substitution attack, suppose that an adversary intercepts a valid pair $[\sv,\tv]$ and inserts $[\sv',\tv']$ with $\sv'\neq\sv$.
Since both pairs are verified under the same key, the substituted pair is accepted if and only if
\begin{equation}
\tv'+\tv
=
\varphi_{\alpha}\bigl((\sv'+\sv)\Gsub\bigr).
\label{eq:substitution-difference}
\end{equation}
In other words, the mask cancels.
Since $\sv'\neq\sv$, the source difference $\sv'+\sv$ is nonzero, and the full row rank of $\Gsub$ ensures that its image is a nonzero codeword of the subcode:
\begin{equation}
    (\sv'+\sv)\Gsub\in \Csub\setminus\{\mathbf{0}_n\}.
\end{equation}

Hence, only nonzero words in $\Csub$ can arise as these differences.
The next proposition, which extends the RM obstruction in \cite[Lem.~2]{liu2024projective}, shows that the exclusion $\mathbf1_n\notin\Csub$ is necessary for $\PS<1$.

\begin{proposition}[All-ones obstruction]
\label{prop:allones}
For each of the three constructions, if $\mathbf{1}_n\in\Csub$, then $\PS=1$.
\end{proposition}
\begin{proof}
Let $\tilde{\sv}$ be a nonzero source difference with $\tilde{\sv}\Gsub=\mathbf1_n$.
Then, for any observed pair $[\sv,\tv]$, the forgery $[\sv+\tilde{\sv},\tv+\mathbf1_{l_t}]$ has a different source and satisfies \eqref{eq:substitution-difference} under every selection, since $\varphi_\alpha(\mathbf1_n)=\mathbf1_{l_t}$.
Thus, the substitution always succeeds.
\end{proof}

\section{Analysis of the Subset Selector}
\label{sec:analysis}
In this section, we analyze the substitution probability of the subset selector.
The reduction and counting arguments below generalize \cite[Thm.~5 and Lems.~6--9]{liu2024projective} to arbitrary binary linear subcodes.
We then distinguish the substitution probability under a fixed coordinate order from a weight-set bound and establish the tightness of this bound over coordinate orders.

\subsection{Reduction of the Substitution Attack}
Let $\tilde{\sv}=\sv'+\sv\neq\mathbf0$ and $\tilde{\tv}=\tv'+\tv$.
Equation~\eqref{eq:substitution-difference} reduces acceptance to $\cv_{\tilde{\sv},\kv_1}=\tilde{\tv}$, where $\cv_{\tilde{\sv}}=\tilde{\sv}\Gsub\in\Csub$ is nonzero.

For $\tilde{\sv}\in\mathcal{S}\setminus\{\mathbf{0}_{l_s}\}$ and $\tilde{\tv}\in\mathcal{T}$, we define
\begin{equation}
N(\cv_{\tilde{\sv}},\tilde{\tv})
:=\bigl|\{\kv_1\in\mathcal{K}_1:
\cv_{\tilde{\sv},\kv_1}=\tilde{\tv}\}\bigr|.
\label{eq:N}
\end{equation}
This quantity counts the selections $\kv_1$ whose projection of $\cv_{\tilde{\sv}}$ equals the specified tag difference.
Then the largest probability of any projection pattern under a uniformly chosen selection $\kv_1$ is given by
\begin{equation}
P_{\max}(\cv_{\tilde{\sv}})
=\max_{\tilde{\tv}\in\mathcal{T}}
\frac{N(\cv_{\tilde{\sv}},\tilde{\tv})}{\binom{n}{l_t}}.
\label{eq:Pt}
\end{equation}
Note that, in general, this probability depends on the coordinate order of $\cv_{\tilde{\sv}}$.

\begin{theorem}[Reduction to projection patterns]
\label{thm:PS-reduce}
For Construction~\ref{con:main}, we have
\begin{equation}
\PSsub=\max_{\tilde{\sv}\neq\mathbf{0}_{l_s}}
P_{\max}(\cv_{\tilde{\sv}}).
\label{eq:PS-reduce}
\end{equation}
\end{theorem}

\begin{proof}
By \eqref{eq:masked-tag-uniform}, observing a valid pair $[\sv,\tv]$ reveals no information about the selection $\kv_1$, and hence all $\binom{n}{l_t}$ selections remain equally likely.
By \eqref{eq:substitution-difference}, a substitution with differences $\tilde{\sv}$ and $\tilde{\tv}$ succeeds for exactly $N(\cv_{\tilde{\sv}},\tilde{\tv})$ of these selections, and its conditional success probability is $N(\cv_{\tilde{\sv}},\tilde{\tv})/\binom{n}{l_t}$.
For each fixed observation $[\sv,\tv]$, every nonzero source difference and every tag difference is realized by choosing $\sv'=\sv+\tilde{\sv}$ and $\tv'=\tv+\tilde{\tv}$.
Then maximizing first over $\tilde{\tv}$ and then over $\tilde{\sv}\neq\mathbf{0}_{l_s}$ gives the right-hand side of \eqref{eq:PS-reduce}.
Note that this optimal conditional success probability is independent of the observed pair.
Consequently, its maximum and its average over valid observations coincide for every source distribution that is independent of the key.
\end{proof}

\subsection{A Weight-Set Bound}
Since evaluating $P_{\max}(\cv_{\tilde{\sv}})$ in Theorem~\ref{thm:PS-reduce} generally requires the coordinate order of $\cv_{\tilde{\sv}}$, we next derive an upper bound that depends only on $w=\wt(\cv_{\tilde{\sv}})$.

To this end, we extend $N(\cv,\tilde{\tv})$ and $P_{\max}(\cv)$ in \eqref{eq:N} and \eqref{eq:Pt} to arbitrary vectors $\cv\in\Ftwo^n$ with the same definitions.
The main idea is to group projection patterns by their Hamming weight.
Note that the number of selections producing a given pattern cannot exceed the number producing patterns of the same weight.
The block vector $[\mathbf{1}_w,\mathbf{0}_{n-w}]$ in the following lemma need not belong to $\Csub$.

\begin{lemma}[Projection-weight count and extremality]
\label{lem:count}
Let $\cv\in\Ftwo^n$ have weight $w$, and let $\tilde{\tv}\in\mathcal{T}$ have weight $w_t$.
\begin{enumerate}
\item[(a)] The number of selections whose projection of $\cv$ has weight $w_t$, summed over all patterns of that weight, is $\binom{w}{w_t}\binom{n-w}{l_t-w_t}$.
Consequently, $N(\cv,\tilde{\tv})\leq\binom{w}{w_t}\binom{n-w}{l_t-w_t}$.
A projection of weight $w_t$ can occur only when
\begin{equation}
\max\{0,\;l_t-(n-w)\}\leq w_t\leq\min\{w,\;l_t\}.
\label{eq:support}
\end{equation}
\item[(b)] For the block vector $\cv^{w}=[\mathbf{1}_w,\mathbf{0}_{n-w}]$ and every $w_t$ in the range \eqref{eq:support}, exactly one projection pattern of weight $w_t$ is possible, namely $[\mathbf{1}_{w_t},\mathbf{0}_{l_t-w_t}]$.
Its count equals the total in (a).
Hence $P_{\max}(\cv)\leq P_{\max}(\cv^{w})$ for every $\cv\in\Ftwo^n$ of weight $w$.
\end{enumerate}
\end{lemma}

\begin{proof}
(a) A projection of weight $w_t$ selects $w_t$ of the $w$ coordinates equal to one and $l_t-w_t$ of the $n-w$ coordinates equal to zero.
Each pair of choices determines a unique selection, which gives the stated product of binomial coefficients.
The count for any one pattern cannot exceed this total.
Also, the feasibility conditions $0\leq w_t\leq w$ and $0\leq l_t-w_t\leq n-w$ give \eqref{eq:support}.
(b) In $\cv^{w}$, all ones precede all zeros.
Since subset projections preserve the coordinate order, every selection taking $w_t$ ones and $l_t-w_t$ zeros produces $[\mathbf{1}_{w_t},\mathbf{0}_{l_t-w_t}]$.
Hence, this single pattern has the entire count in (a).
Then normalizing by $\binom{n}{l_t}$ and maximizing over the feasible values of $w_t$ proves the inequality.
\end{proof}

For each integer $0\leq w\leq n$, we define the largest projection-weight probability as
\begin{equation}
H_{n,l_t}(w)
:=
\max_{w_t}
\frac{\binom{w}{w_t}\binom{n-w}{l_t-w_t}}
{\binom{n}{l_t}},
\label{eq:H}
\end{equation}
where the maximum is taken over the integer range \eqref{eq:support}.
Then, for every $\cv\in\Ftwo^n$ of weight $w$, Lemma~\ref{lem:count} gives $P_{\max}(\cv)\leq H_{n,l_t}(w)=P_{\max}(\cv^{w}).$
In other words, $H_{n,l_t}(w)$ is the exact maximum of $P_{\max}(\cv)$ over all binary vectors of length $n$ and weight $w$.
Note that $P_{\max}(\cv)$ is the largest probability of an individual projection pattern of $\cv$, while $H_{n,l_t}(w)$ is the largest probability of a projection weight.
Maximizing the latter over the nonzero weights of the subcode gives the weight-set bound $B(\Csub)$ defined below.

\begin{lemma}[Hypergeometric mode]
\label{lem:mode}
Let $W_t$ denote the weight of the projection of a fixed vector of weight $w$ under a uniformly chosen selection $\kv_1$.
Then $W_t\sim\mathrm{Hypergeometric}(n,w,l_t)$, and the ratio maximized in \eqref{eq:H} equals $\Pr[W_t=w_t]$.
Its maximum occurs at
\begin{equation}
w_t^{\star}
=
\Bigl\lfloor\frac{(w+1)(l_t+1)}{n+2}\Bigr\rfloor.
\label{eq:mode}
\end{equation}
If the fraction in \eqref{eq:mode} is an integer, both $w_t^{\star}-1$ and $w_t^{\star}$ are modes.
Otherwise, $w_t^{\star}$ is the unique mode.
Furthermore, $H_{n,l_t}(w)=H_{n,l_t}(n-w)$.
\end{lemma}

\begin{proof}
Note that Lemma~\ref{lem:count}(a) gives the hypergeometric mass function $p(j)$.
For consecutive feasible values, $p(j+1)/p(j)=(w-j)(l_t-j)/[(j+1)(n-w-l_t+j+1)]$, which is at least one exactly when $j+1\leq(w+1)(l_t+1)/(n+2)$, proving the mode formula \cite{ross2019first}.
Replacing $(w,j)$ by $(n-w,l_t-j)$ interchanges the binomial factors and proves the symmetry.
\end{proof}

Since $\Gsub$ has full row rank, nonzero source differences correspond bijectively to nonzero codewords of $\Csub$.
Then we define the nonzero weight set of $\Csub$ and its associated \emph{weight-set bound} as follows:
\begin{equation}
    B(\Csub):=\max_{w\in\mathcal{W}(\Csub)}H_{n,l_t}(w),
    \label{eq:Wset}
\end{equation}
where 
\begin{equation}
\mathcal{W}(\Csub)
:=
\{\wt(\cv_{\tilde{\sv}}):
  \tilde{\sv}\in\mathcal{S}\setminus\{\mathbf{0}_{l_s}\}\}.
  \label{eq:Wset_weight}
\end{equation}

\begin{corollary}[Weight-set bound]
\label{cor:cert}
For Construction~\ref{con:main}, the substitution probability satisfies
\begin{equation}
\PSsub\leq B(\Csub).
\end{equation}
\end{corollary}

\begin{proof}
By Theorem~\ref{thm:PS-reduce} and Lemma~\ref{lem:count}, we have
\[
\begin{aligned}
\PSsub&=\max_{\cv\in\Csub\setminus\{\mathbf0\}}P_{\max}(\cv)\\
&\leq\max_{w\in\mathcal W(\Csub)}H_{n,l_t}(w)=B(\Csub).
\end{aligned}
\]
\end{proof}

Note that $B(\Csub)\geq1/(l_t+1)$, since each $H_{n,l_t}(w)$ is the largest of the probabilities of at most $l_t+1$ projection weights in \eqref{eq:H}, which sum to one.
Also, Lemma~\ref{lem:mode} evaluates $H_{n,l_t}$ once the weights are known, while obtaining those weights remains a separate enumeration problem for each subcode.

\subsection{Interpreting the Weight-Set Bound}

The subset selector reads the selected bits in increasing coordinate index, and hence its substitution probability depends on the coordinate order, i.e., the column order of $\Gsub$.
Reordering the columns preserves the nonzero weight set, and hence also the weight-set bound.
The following example shows that this bound can be strict for a given coordinate order.
We then prove that some coordinate order meets it.

\begin{example}[Same weights, different substitution probabilities]
\label{ex:ordering-tightness}
Let $l_s=1$, $n=4$, $l_t=2$, and $\Csub=\{\mathbf{0}_4,[0,1,0,1]\}$.
Among the six selections, the projection $[0,1]$ of the only nonzero codeword occurs three times, while each other two-bit pattern occurs once.
Hence, $\PSsub(\Csub)=3/6=1/2$.

Now, suppose that the coordinate order is changed to $(2,4,1,3)$.
Then the nonzero codeword becomes $[1,1,0,0]$, whose projection $[1,0]$ occurs under four of the six selections, and the substitution probability for this coordinate order is $4/6=2/3$.

Note that both coordinate orders have the same nonzero weight set $\{2\}$ and the same bound $B(\Csub)=H_{4,2}(2)=2/3$.
In other words, the bound is strict for the original coordinate order and is met after reordering.
\end{example}

For a coordinate permutation $\pi$ of $[n]$, we define
\[
\pi\cv=[c_{\pi(1)},\ldots,c_{\pi(n)}],
\qquad
\pi\Csub=\{\pi\cv:\cv\in\Csub\}.
\]
Note that permuting the columns of $\Gsub$ preserves all codeword weights, but it can change the projection-pattern counts under the coordinate order of the subset selector.

\begin{corollary}[Tightness of the weight-set bound]
\label{thm:minimax}
For every subcode $\Csub$,
\begin{equation}
\max_{\pi}\PSsub(\pi\Csub)=B(\Csub),
\label{eq:minimax}
\end{equation}
where the maximum is over all coordinate permutations.
\end{corollary}

\begin{proof}
Since coordinate permutations preserve the weight set, we have $\PSsub(\pi\Csub)\leq B(\Csub)$ by Corollary~\ref{cor:cert}.
Now, choose a nonzero codeword of weight $w$ with $H_{n,l_t}(w)=B(\Csub)$, and permute its ones before its zeros.
Then Lemma~\ref{lem:count}(b) gives a projection pattern with probability $B(\Csub)$, and hence equality holds.
\end{proof}

Corollary~\ref{thm:minimax} implies that no upper bound that depends only on the nonzero weight set and holds for every coordinate order can be smaller than $B(\Csub)$.
For a fixed coordinate order, equality holds in particular when the subcode contains $[\mathbf1_w,\mathbf0_{n-w}]$ at a weight maximizing $H_{n,l_t}(w)$ over its nonzero weight set.
For instance, the first-order RM subcode with the canonical order in Section~\ref{sec:rm} has this property.
This explains when the expression of \cite{liu2024projective}, which assumes a block vector, is exact.
Otherwise, it is only an upper bound.

\section{Analysis of the Ordered and Independent Selectors}
\label{sec:ordered}

In Section~\ref{sec:analysis}, we showed that the substitution probability of the subset selector depends on the order of the codeword coordinates.
In this section, we analyze the other two selectors.
We first establish the exact substitution probability of the ordered selector and its relation to the bounds for the subset selector.
Then we analyze the independent selector and compare the two selectors.

Throughout this section, we assume $1\leq l_t<n$ and $\mathbf1_n\notin\Csub$.
We define the \emph{maximum imbalance} of the subcode as
\begin{equation}
\rho(\Csub):=\max_{w\in\mathcal W(\Csub)}\frac{|2w-n|}{n}\in[0,1).
\label{eq:rho}
\end{equation}
Then $\rho(\Csub)\leq\epsilon$ if and only if every nonzero codeword has relative weight in $[(1-\epsilon)/2,(1+\epsilon)/2]$, i.e., if and only if $\Csub$ is $\epsilon$-balanced \cite{naor1993small,tashma2017explicit}.
In particular, exact balance means $\rho(\Csub)=0$, or equivalently, every nonzero codeword has weight $n/2$.
We call $\Csub$ balanced if $\rho(\Csub)=0$ and unbalanced otherwise.
In the rest of this section, we write $\rho=\rho(\Csub)$.

\subsection{Substitution Probability of the Ordered Selector}
\label{ssec:ordered-security}

In Construction~\ref{con:ordered}, the selection is uniform over the $(n)_{l_t}$ ordered tuples of distinct coordinates.
Hence, for a fixed codeword, all projection patterns of the same weight are equally likely.
With the falling factorial from Section~\ref{ssec:con-ordered}, we define
\begin{equation}
R_{n,l_t}(w):=\max_{w_t}\frac{(w)_{w_t}\,(n-w)_{l_t-w_t}}{(n)_{l_t}},
\label{eq:R}
\end{equation}
where the maximum is over the integer range \eqref{eq:support}.
The following theorem shows that this quantity is the largest probability of any projection pattern of a weight-$w$ codeword under the ordered selector.

\begin{theorem}[Deception probabilities of the ordered selector]
\label{thm:ordered}
For Construction~\ref{con:ordered}, we have $\PI=2^{-l_t}$ and
\begin{equation}
P_{S,\mathrm{ord}}=\max_{w\in\mathcal{W}(\Csub)}R_{n,l_t}(w).
\label{eq:ordered}
\end{equation}
These probabilities are invariant under every coordinate permutation of $\Csub$.
\end{theorem}

\begin{proof}
Theorem~\ref{thm:PI} gives the impersonation probability.
Since the ordered extraction map is linear, the reduction in the proof of Theorem~\ref{thm:PS-reduce} applies.
In other words, $\PS$ is the largest probability of any projection pattern over all nonzero codewords of the subcode.

Now, fix a nonzero codeword of weight $w$ and a tag $\tilde{\tv}$ of weight $w_t$.
An ordered sample produces $\tilde{\tv}$ exactly when two conditions hold.
First, the tag positions where $\tilde{\tv}$ equals one receive distinct coordinates at which the codeword equals one.
Second, the remaining positions receive distinct coordinates at which the codeword equals zero.
There are $(w)_{w_t}$ choices for the former and $(n-w)_{l_t-w_t}$ for the latter.
Since these choices concern disjoint coordinate sets, their product is the number of ordered samples producing the tag.
Then dividing by $(n)_{l_t}$ and maximizing over the feasible tag weights and the nonzero codeword weights gives \eqref{eq:ordered}.
Since the expression depends on the codeword only through its weight, coordinate permutations leave these probabilities unchanged.
\end{proof}

Theorem~\ref{thm:ordered} expresses the substitution probability as a maximum over the nonzero codeword weights.
The next proposition reduces it to the single parameter $\rho$.

\begin{proposition}[Reduction to maximum imbalance]
\label{prop:ordered-imbalance}
We have
\begin{equation}
\PSord=
\begin{cases}
\dfrac{\binom{n(1+\rho)/2}{l_t}}{\binom n{l_t}},&l_t\leq n\rho,\\[8pt]
\dfrac{\binom{n-l_t}{\lfloor(n-l_t)/2\rfloor}}{\binom n{n(1-\rho)/2}},&l_t>n\rho.
\end{cases}
\label{eq:ordered-imbalance-exact}
\end{equation}
For fixed $n$ and $l_t$, $\PSord$ depends on $\Csub$ only through $\rho$ and increases strictly with $\rho$ over its admissible values.
\end{proposition}

Here, $n(1+\rho)/2$ and $n(1-\rho)/2$ are the numbers of majority and minority bits in a most imbalanced nonzero codeword.
For $l_t\leq n\rho$, \eqref{eq:ordered-imbalance-exact} is the probability that every selected coordinate of this codeword holds a majority bit.
Otherwise, a most likely pattern leaves the unselected bits as balanced as possible.

\begin{proof}
Fix a nonzero codeword of weight $w$.
By the proof of Theorem~\ref{thm:ordered}, a pattern of weight $w_t$ has probability
\begin{equation}
\frac{(w)_{w_t}(n-w)_{l_t-w_t}}{(n)_{l_t}}
=\frac{\binom{n-l_t}{w-w_t}}{\binom nw},
\end{equation}
where the numerator counts the placements of the remaining ones among the unselected coordinates.
Complementing both the codeword and the pattern leaves this probability unchanged.
Thus, $R_{n,l_t}(w)=R_{n,l_t}(n-w)$, and we may assume $w\leq n/2$.
Then $w-w_t$ ranges over the integers from $\max\{0,w-l_t\}$ to $\min\{w,n-l_t\}$, and the numerator is maximized at $w-w_t=\min\{w,\lfloor(n-l_t)/2\rfloor\}$, since the binomial coefficients increase toward the center.
This value is feasible, since it is at most $\min\{w,n-l_t\}$ and $w-l_t\leq n/2-l_t\leq(n-l_t)/2$.
Hence, for $w\leq n/2$, we have
\begin{equation}
R_{n,l_t}(w)=\frac{\binom{n-l_t}{\min\{w,\lfloor(n-l_t)/2\rfloor\}}}{\binom nw}.
\label{eq:R-closed}
\end{equation}

Furthermore, for integers $0\leq w<\lfloor n/2\rfloor$, the ratio $R_{n,l_t}(w+1)/R_{n,l_t}(w)$ equals $(n-l_t-w)/(n-w)$ if $w<\lfloor(n-l_t)/2\rfloor$ and $(w+1)/(n-w)$ otherwise.
Both ratios are below one because $l_t\geq1$ and $2w+1<n$.
Therefore, $R_{n,l_t}(w)$ decreases strictly as $\min\{w,n-w\}$ grows.
By \eqref{eq:rho}, the smallest value of $\min\{w,n-w\}$ over the nonzero codewords is $n(1-\rho)/2$.
Then Theorem~\ref{thm:ordered} gives $\PSord=R_{n,l_t}(n(1-\rho)/2)$, which increases strictly with $\rho$.
Finally, $w=n(1-\rho)/2$ satisfies $w\leq\lfloor(n-l_t)/2\rfloor$ if and only if $l_t\leq n\rho$, and in this case, \eqref{eq:R-closed} equals $\binom{n-w}{l_t}/\binom n{l_t}$, which gives \eqref{eq:ordered-imbalance-exact}.
\end{proof}

Having determined $\PSord$, we next compare it with the substitution probability of the subset selector for the same subcode.
The following bounds hold for every codeword and therefore also after maximizing over the subcode.

\begin{theorem}[Weight-dependent bounds]
\label{thm:sandwich}
For every $\cv\in\Ftwo^n$ with $\wt(\cv)=w$, we have
\begin{equation}
R_{n,l_t}(w)\leq P_{\max}(\cv)\leq H_{n,l_t}(w).
\end{equation}
Consequently, the ordered selector in Construction~\ref{con:ordered} and the subset selector, using the same subcode $\Csub$, satisfy
\begin{equation}
\begin{aligned}
\PSord(\Csub)
&=\max_{w\in\mathcal{W}(\Csub)}R_{n,l_t}(w)\\
&\leq\PSsub(\Csub)\leq B(\Csub).
\end{aligned}
\end{equation}
Equality holds in the upper bound for $\cv=\cv^{w}$.
\end{theorem}

\begin{proof}
The upper bound and the equality for $\cv^{w}$ follow from Lemma~\ref{lem:count} and the definition in \eqref{eq:H}.
For the lower bound, fix a feasible $w_t$.
Then Lemma~\ref{lem:count}(a) counts $\binom{w}{w_t}\binom{n-w}{l_t-w_t}$ selections $\kv_1$ whose projection has weight $w_t$.
Note that the selection counts over all $\binom{l_t}{w_t}$ possible patterns of weight $w_t$ sum to $\binom{w}{w_t}\binom{n-w}{l_t-w_t}$.
Therefore, at least one pattern has a count no smaller than this sum divided by $\binom{l_t}{w_t}$.
Consequently, we have
\begin{equation}
\begin{aligned}
    P_{\max}(\cv)
&\geq\max_{w_t}
\frac{\binom{w}{w_t}\binom{n-w}{l_t-w_t}}
{\binom{n}{l_t}\binom{l_t}{w_t}}\\
&=\max_{w_t}\frac{(w)_{w_t}(n-w)_{l_t-w_t}}{(n)_{l_t}}
=R_{n,l_t}(w).
\end{aligned}
\end{equation}
Maximizing the pointwise bounds over all nonzero codeword differences gives $\PSord\leq\PSsub\leq B(\Csub)$.
Here, the three maxima are $\PSord$ by Theorem~\ref{thm:ordered}, $\PSsub$ by Theorem~\ref{thm:PS-reduce}, and $B(\Csub)$ by definition.
\end{proof}

\subsection{Substitution Probability of the Independent Selector}
\label{ssec:iid}

The ordered selector removes the dependence on the coordinate order, but selecting distinct coordinates still makes the sampled bits dependent.
In contrast, in Construction~\ref{con:iid}, each coordinate index is chosen independently and uniformly, allowing repetitions.

\begin{theorem}[Deception probabilities of the independent selector]
\label{thm:iid}
For Construction~\ref{con:iid}, we have
\begin{equation}
\begin{aligned}
\PI&=2^{-l_t},\\
\PSind&=\max_{w\in\mathcal{W}(\Csub)}\Bigl(\frac{\max\{w,n-w\}}{n}\Bigr)^{l_t}=\Bigl(\frac{1+\rho}{2}\Bigr)^{l_t}.
\end{aligned}
\label{eq:iid}
\end{equation}
These probabilities are invariant under every coordinate permutation of $\Csub$.
For fixed $n$ and $l_t$, $\PSind$ increases strictly with $\rho$ over its admissible values.
\end{theorem}

\begin{proof}
Theorem~\ref{thm:PI} gives $\PI$.
Since uniform masking leaves the selection uniform after an observation, the substitution reduces to a nonzero codeword difference as in Theorem~\ref{thm:PS-reduce}.
For a difference of weight $w$, a single uniformly chosen coordinate takes its more frequent value with probability $\max\{w,n-w\}/n$.
Then independence across the $l_t$ positions makes the largest pattern probability equal to the $l_t$th power of this value.
The pattern whose bits all take the majority value has this probability.
Maximizing over $w\in\mathcal W(\Csub)$ proves \eqref{eq:iid} and the invariance under coordinate permutations.
Finally, the explicit formula is strictly increasing in $\rho$.
\end{proof}

Note that, for a balanced nonzero codeword, independent sampling produces independent unbiased bits.
Thus, its unmasked projection is uniform over $\mathcal T$, which leads to the following characterization.

\begin{corollary}[Characterization of the optimal independent selector]
\label{cor:optimal}
The independent selector satisfies $\PSind=\PI=2^{-l_t}=|\mathcal T|^{-1}$ if and only if $\rho(\Csub)=0$, that is, every nonzero codeword has weight $n/2$.
\end{corollary}

\subsection{Comparison of the Ordered and Independent Selectors}
\label{ssec:ord-ind-comparison}

Proposition~\ref{prop:ordered-imbalance} and Theorem~\ref{thm:iid} show that a most imbalanced nonzero codeword determines both substitution probabilities, namely $\PSord$ in \eqref{eq:ordered-imbalance-exact} and $\PSind$ in \eqref{eq:iid}.
However, their relative sizes depend on the parameters, and the following corollary, proved in Appendix~\ref{app:ord-ind-proof}, compares them.

\begin{corollary}[Comparison of the ordered and independent selectors]
\label{cor:ord-ind-comparison}
For the same subcode, with $\mathbf1_n\notin\Csub$, and the same integer tag length $1\leq l_t<n$, the following hold:
\begin{enumerate}
\item If $l_t=1$, then $\PSord=\PSind=(1+\rho)/2$.
\item If $2\leq l_t<n$ and $\rho=0$, then $\PSind=2^{-l_t}<\PSord$.
\item If $2\leq l_t\leq\lfloor n/2\rfloor$ and $\rho>0$, then $\PSord<\PSind$.
\end{enumerate}
Items~2 and~3 give, for $2\leq l_t\leq\lfloor n/2\rfloor$,
\begin{equation}
\begin{cases}
\PSind<\PSord,&\rho=0,\\
\PSord<\PSind,&\rho>0.
\end{cases}
\label{eq:half-block-comparison}
\end{equation}
\end{corollary}
\begin{figure}[t]
\centering
\includegraphics{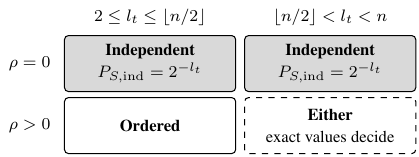}
\caption{The selector with the smaller substitution probability for the same subcode and tag length $2\leq l_t<n$ (Corollary~\ref{cor:ord-ind-comparison}).
For $l_t=1$, both selectors have $\PS=(1+\rho)/2$.}
\label{fig:ord-ind}
\end{figure}

For unbalanced subcodes with $l_t>\lfloor n/2\rfloor$, the comparison depends on the exact values in \eqref{eq:ordered-imbalance-exact} and \eqref{eq:iid}.
In this case, either selector can have the smaller substitution probability.
Fig.~\ref{fig:ord-ind} summarizes these cases.

For balanced subcodes and fixed $l_t$, $\PSord$ approaches $2^{-l_t}$ as $n\to\infty$ over blocklengths that admit balanced subcodes.
In contrast, the independent selector achieves this value at every such blocklength.

\begin{remark}[Neither the subset nor the ordered selector achieves $2^{-l_t}$]
\label{rem:no-optimum}
For $2\leq l_t<n$, neither the subset nor the ordered selector achieves $2^{-l_t}$ for any subcode.
To see this, fix a nonzero codeword and build a projection pattern bit by bit, each time choosing the value held by more of the unselected coordinates, either value on a tie.
These counts depend only on the values chosen so far, and hence this pattern is fixed.
Under the ordered selector, each draw produces the chosen value with probability at least $1/2$, and above $1/2$ when the number of unselected coordinates is odd.
Since $l_t\geq2$, the first two draws see $n$ and $n-1$ unselected coordinates, one of which is odd.
Hence, this pattern has probability above $2^{-l_t}$, and thus $\PSord>2^{-l_t}$ and $\PSsub>2^{-l_t}$ by Theorem~\ref{thm:sandwich}.
\end{remark}

\begin{table*}[t]
\centering
\caption{Selector comparison.
$L_\alpha$ is the selector entropy, and the key length is $\lceil L_\alpha\rceil+l_t$, including an independent uniform $l_t$-bit mask.
Bounds apply to the subset selector, and equalities to the other two.
Here $R=\PSord$ is given by \eqref{eq:ordered-imbalance-exact}, and $B\geq1/(l_t+1)$.}
\label{tab:baseline}
\footnotesize
\setlength{\tabcolsep}{6pt}
\begin{tabular}{lccccc}
\toprule
Selector & Selector entropy $L_\alpha$ (bits) & $\PI$ & $\PS$ & Mask bits & Depends on order\\
\midrule
Subset (Construction~\ref{con:main}) & $\log_2\binom n{l_t}$ & $2^{-l_t}$ & $R\leq\PS\leq B$ & $l_t$ & yes\\
Ordered (Construction~\ref{con:ordered}) & $\log_2(n)_{l_t}$ & $2^{-l_t}$ & $R$ & $l_t$ & no\\
Independent (Construction~\ref{con:iid}) & $l_t\log_2 n$ & $2^{-l_t}$ & $((1+\rho)/2)^{l_t}$ & $l_t$ & no\\
\bottomrule
\end{tabular}
\end{table*}

\subsection{Comparison of Key Lengths}
\label{ssec:baseline}

Table~\ref{tab:baseline} lists the selector entropy $L_\alpha$ and the mask length of each construction.
Hence, the key length is $\lceil L_\alpha\rceil+l_t$ bits, as in \eqref{eq:keylen}, \eqref{eq:keylen-ordered}, and \eqref{eq:keylen-independent}.
Furthermore, the selector entropies satisfy
\begin{equation}
\log_2\tbinom{n}{l_t}\;\leq\;\log_2(n)_{l_t}\;\leq\;l_t\log_2 n,
\label{eq:keyorder}
\end{equation}
with both inequalities strict for $l_t\geq2$ and equality throughout for $l_t=1$.
Also, the difference between the selector entropies of the ordered and independent selectors tends to zero as $n\to\infty$ for fixed $l_t$.

For instance, consider $n=1024$ and the target $\PS\leq2^{-64}$.
Then a balanced subcode requires $l_t=65$ for the ordered selector and $l_t=64$ for the independent selector.
The corresponding key lengths are $648+65=713$ and $640+64=704$ bits, respectively.
At $\rho=1/16$, the tag lengths become $67$ and $71$, and the key lengths are $667+67=734$ and $710+71=781$ bits.
As a result, balance affects both which selector has the smaller substitution probability and the key length.

\begin{remark}[Selected-column implementation]
\label{rem:matrix-implementation}
Let $\mathbf g_i$ be the $i$th column of $\Gsub$.
Every fixed selection $\alpha$ induces $\mathbf A_\alpha=[\mathbf g_{i_1},\ldots,\mathbf g_{i_{l_t}}] \in\Ftwo^{l_s\times l_t}$ in its selected order, and $\varphi_\alpha(\sv\Gsub)=\sv\mathbf A_\alpha$.
Hence, a party can store the selected coordinate indices and compute only the corresponding codeword bits as needed, without generating the full codeword.
Alternatively, it can precompute the selected-column matrix $\mathbf A_\alpha$, which occupies $l_s l_t$ bits of working storage.
This working storage is separate from the key.
\end{remark}

\section{Balanced and Near-Balanced Subcodes}
\label{sec:rm}

In this section, we study subcodes with a small maximum imbalance.
Recall that, for fixed $n$ and $l_t$, exact balance minimizes both the ordered and the independent substitution probabilities (Proposition~\ref{prop:ordered-imbalance} and Theorem~\ref{thm:iid}).
Also, it is necessary and sufficient for $\PSind=2^{-l_t}$ (Corollary~\ref{cor:optimal}).
We first recall that the first-order RM subcode is exactly balanced and establish the limit that exact balance imposes on the source dimension.
Then we relax balance to support longer sources.

\subsection{The First-Order RM Subcode}

\begin{definition}[First-order RM subcode]
\label{def:rm}
For $m\geq1$ and $n=2^m$, let $\mathbf G_1\in\Ftwo^{m\times n}$ contain every binary $m$-vector as a column, and set $\Csub^{\circ}=\{\uv\mathbf G_1:\uv\in\Ftwo^m\}$.
It is the subcode of the first-order RM code $\mathrm{RM}(m,1)$ obtained by dropping the all-ones vector from its standard basis \cite{macwilliams1977theory,abbe2020reed}.
In other words, the all-ones row of its standard generator matrix is deleted.
In the canonical order, the first row is $[\mathbf1_{n/2},\mathbf0_{n/2}]$.
Appendix~\ref{app:repro} specifies this order completely.
\end{definition}

Every nonzero codeword of $\Csub^{\circ}$ has weight $n/2$ \cite{macwilliams1977theory,abbe2020reed}.
Thus, $\mathbf1_n\notin\Csub^{\circ}$, $\mathcal W(\Csub^{\circ})=\{n/2\}$, and $\rho(\Csub^{\circ})=0$.
The source dimension is $l_s=m$.
Also, selecting fewer generator rows gives a balanced subcode with a correspondingly smaller source space.

\begin{remark}[Relation to uniform linear hashing]
\label{rem:uniform-matrix}
For $\Csub^{\circ}$, and also when only $l_s<m$ of its generator rows are retained, every $l_s$-bit vector occurs equally often as a column of $\Gsub$.
Hence, independent sampling selects independent uniform columns, and $\mathbf A_\alpha$ in Remark~\ref{rem:matrix-implementation} is a uniformly random binary matrix.
In other words, in the balanced case, the independent selector on these subcodes realizes the uniform linear hash $\sv\mapsto\sv\mathbf A_\alpha$ followed by a mask.
Such a masked linear hash is a classical universal-hashing construction \cite{wegman1981new}.
Here, we compute it with an existing channel encoder.
Moreover, we represent this hash through codeword coordinates.
This places it in the same framework as the subset and ordered selectors on the same code.
\end{remark}

\begin{corollary}[Substitution probabilities for the first-order RM subcode]
\label{thm:rm}
Under the canonical order in Definition~\ref{def:rm},
\begin{equation}
\begin{aligned}
\PSsub(\Csub^{\circ})&=H_{n,l_t}(n/2),\\
\PSord(\Csub^{\circ})&=R_{n,l_t}(n/2),\\
\PSind(\Csub^{\circ})&=2^{-l_t}.
\end{aligned}
\label{eq:rmexact}
\end{equation}
The last two equalities hold for every coordinate order.
\end{corollary}

\begin{proof}
By Lemma~\ref{lem:count}(b), the first row has $P_{\max}=H_{n,l_t}(n/2)$, which also bounds $P_{\max}$ for every nonzero codeword.
Then the subset equality follows from Theorem~\ref{thm:PS-reduce}, and the other equalities follow from Theorems~\ref{thm:ordered} and \ref{thm:iid}.
\end{proof}

For $w=n/2$, Lemma~\ref{lem:mode} places the maximum in $H_{n,l_t}(n/2)$ at $w_t\in\{\lfloor l_t/2\rfloor,\lceil l_t/2\rceil\}$.
Also, in \eqref{eq:R}, the ratio of consecutive feasible terms is $(n/2-w_t)/(n/2-l_t+w_t+1)$, and hence $R_{n,l_t}(n/2)$ is maximized at the same tag weights.
Consequently, we have
\begin{equation}
\begin{aligned}
R_{n,l_t}(n/2)
&=\frac{(n/2)_{\lfloor l_t/2\rfloor}(n/2)_{\lceil l_t/2\rceil}}
{(n)_{l_t}},\\
H_{n,l_t}(n/2)
&=\binom{l_t}{\lfloor l_t/2\rfloor}R_{n,l_t}(n/2).
\end{aligned}
\label{eq:balanced-explicit}
\end{equation}
Note that the canonical order is the worst coordinate order for the subset selector on $\Csub^{\circ}$.
Indeed, $\mathcal W(\Csub^{\circ})=\{n/2\}$ gives $B(\Csub^{\circ})=H_{n,l_t}(n/2)$, which equals $\PSsub(\Csub^{\circ})$ by \eqref{eq:rmexact} and is the largest value over all coordinate orders by Corollary~\ref{thm:minimax}.
By \eqref{eq:balanced-explicit}, this value is $\binom{l_t}{\lfloor l_t/2\rfloor}$ times $\PSord(\Csub^{\circ})$.
The reason is that the subset selector always reads the first row $[\mathbf1_{n/2},\mathbf0_{n/2}]$ as $[\mathbf1_{w_t},\mathbf0_{l_t-w_t}]$, while the ordered selector spreads the same probability over all $\binom{l_t}{w_t}$ arrangements of these bits.

\begin{proposition}[Source dimension under exact balance]
\label{prop:balanced}
If every nonzero codeword of a binary linear $(n,l_s)$ code has weight $n/2$, then $l_s\leq\log_2 n$.
At $n=2^m$, the subcode $\Csub^{\circ}$ has the largest possible source dimension $l_s=m$.
\end{proposition}

\begin{proof}
Let $\mathbf g_1,\ldots,\mathbf g_n$ be the columns of a generator matrix $\mathbf G$ of the code, so that the $j$th bit of $\sv\mathbf G$ is $\sv\mathbf g_j$.
We first count, for a nonzero $\mathbf g\in\Ftwo^{l_s}$, the vectors $\sv\in\Ftwo^{l_s}$ with $\sv\mathbf g=1$.
Since $\mathbf g\neq\mathbf0$, it has a one at some position $i$.
Let $\mathbf e_i$ be the unit vector with a one at position $i$, so that $\mathbf e_i\mathbf g$ equals the $i$th entry of $\mathbf g$, i.e., $\mathbf e_i\mathbf g=1$.
For each $\sv$, the vector $\sv+\mathbf e_i$ is $\sv$ with its $i$th bit flipped.
Hence, $\Ftwo^{l_s}$ is partitioned into the $2^{l_s-1}$ two-element sets $\{\sv,\sv+\mathbf e_i\}$, one for each $\sv$ whose $i$th bit is zero.
Within each set, linearity gives $(\sv+\mathbf e_i)\mathbf g=\sv\mathbf g+1$, so one vector gives $\sv\mathbf g=0$ and the other gives $\sv\mathbf g=1$.
Thus, exactly one vector in each set satisfies $\sv\mathbf g=1$, and therefore exactly $2^{l_s-1}$ vectors $\sv$ do.
Summing over the columns, the total weight of all codewords is $(n-z)2^{l_s-1}$, where $z$ is the number of zero columns.
Counting by codewords instead, the zero codeword contributes nothing, and each of the other $2^{l_s}-1$ codewords contributes $n/2$ by assumption, and hence the same total equals $(2^{l_s}-1)n/2$.
Equating the two totals gives $(n-z)2^{l_s-1}=(2^{l_s}-1)n/2$, which simplifies to $n=z2^{l_s}$.
Since $n\geq1$, the integer $z$ is positive, and hence $n\geq2^{l_s}$, i.e., $l_s\leq\log_2 n$.
Finally, $\Csub^{\circ}$ meets this bound at $n=2^m$.
\end{proof}

Note that the dimension bound applies to exactly balanced codes at a fixed blocklength $n$.
Hence, when the coordinates come from one code, supporting $l_s>\log_2 n$ requires departing from exact balance.
Next, we quantify the resulting increase in both substitution probabilities.

\subsection{Larger Source Dimensions Through Near Balance}
\label{ssec:nearbalanced}

With $\rho=\rho(\Csub)$, Theorem~\ref{thm:iid} gives
\begin{equation}
\PSind=2^{-l_t}(1+\rho)^{l_t}.
\label{eq:rhopenalty}
\end{equation}
Thus, imbalance increases $\PSind$ by the factor $(1+\rho)^{l_t}$ relative to an exactly balanced subcode.
Since $0\leq\rho<1$ by \eqref{eq:rho}, the necessary and sufficient condition for $\PSind\leq2^{-\lambda}$, with $\lambda>0$, is given by
\begin{equation}
l_t\geq
\left\lceil
\frac{\lambda}{1-\log_2(1+\rho)}
\right\rceil.
\label{eq:rho-taglength}
\end{equation}
Within the comparison range $1\leq l_t<n$, this target is feasible if and only if the right-hand side of \eqref{eq:rho-taglength} is at most $n-1$.

Proposition~\ref{prop:balanced} limits exact balance to $l_s\leq\log_2 n$, and hence longer sources need near-balanced subcodes, i.e., subcodes with a small $\rho$ that can be bounded without enumerating all $2^{l_s}-1$ nonzero weights.
For this purpose, we use the extended BCH-dual code, i.e., the dual of the length-$(n-1)$ primitive narrow-sense binary BCH code of designed distance $2t+1$, extended by an identically zero coordinate.
We choose this code for four reasons.
First, the Carlitz--Uchiyama bound \cite[Ch.~9, Thm.~18]{macwilliams1977theory} keeps every nonzero weight within $(t-1)\sqrt n$ of $n/2$, which certifies $\rho\leq(2t-2)/\sqrt n$ and $\mathbf1_n\notin\Csub$ for every subcode (Corollary~\ref{prop:bch-guarantee}).
Second, for $2t-1<\sqrt n$, its dimension is $mt$ \cite[Ch.~9, Cor.~8]{macwilliams1977theory}, and hence $t$ trades the source dimension against this bound.
Third, it contains the first-order RM subcode, since the first $m$ rows of the generator in Appendix~\ref{app:repro} generate a coordinate permutation of this subcode.
Hence, its row prefixes are exactly balanced up to $l_s=m$ and extend to $l_s=mt$.
Fourth, the code is cyclic before the extension, and hence, similar to other cyclic codes, it can be encoded by a linear-feedback shift register \cite{macwilliams1977theory}.
The following corollary bounds both substitution probabilities without enumerating weights.

\begin{table}[t]
\centering
\caption{Exact values of $\PSsub$ under two coordinate orders at $(n,l_s,l_t)=(128,7,16)$.
Both subcodes are balanced.}
\label{tab:orderings}
\footnotesize
\setlength{\tabcolsep}{10pt}
\begin{tabular}{lc}
\toprule
Coordinate order & $\PSsub$\\
\midrule
Canonical (RM) & $0.20988$\\
Cyclic (BCH-dual) & $1.4527\times10^{-4}$\\
\bottomrule
\end{tabular}
\end{table}

\begin{corollary}[BCH-dual subcodes]
\label{prop:bch-guarantee}
Let $m\geq2$ and $n=2^m$ with $2t-1<\sqrt n$, and let $\Csub$ be a subcode of the extended BCH-dual code above.
Then $\mathbf1_n\notin\Csub$ and $\rho\leq\bar\rho:=(2t-2)/\sqrt n<1$.
Moreover, with $b_\star:=n/2-\lfloor(t-1)\sqrt n\rfloor$, every $1\leq l_t<n$ satisfies
\begin{equation}
\PSord\leq R_{n,l_t}(b_\star),
\qquad
\PSind\leq(1-b_\star/n)^{l_t}.
\label{eq:bch-bounds}
\end{equation}
If also $t\geq2$ and $l_t\leq\lfloor n/2\rfloor$, then the bounds in \eqref{eq:bch-bounds} satisfy
\begin{equation}
2^{-l_t}\leq R_{n,l_t}(b_\star)\leq(1-b_\star/n)^{l_t}\leq2^{-l_t}(1+\bar\rho)^{l_t}.
\label{eq:bch-gap}
\end{equation}
\end{corollary}

\begin{proof}
Note that the appended zero coordinate does not change any weight, and that $2t-1<\sqrt n\leq2^{\lceil m/2\rceil}$.
Hence, the Carlitz--Uchiyama bound \cite[Ch.~9, Thm.~18]{macwilliams1977theory} gives $|w-n/2|\leq(t-1)\sqrt n<n/2$ for every nonzero weight $w$ of $\Csub$.
Then $\mathbf1_n\notin\Csub$ and $\rho\leq\bar\rho<1$, and by integrality, every nonzero codeword has at least $b_\star$ minority bits, which gives \eqref{eq:bch-bounds} by the monotonicity in Proposition~\ref{prop:ordered-imbalance} and Theorem~\ref{thm:iid}.

For \eqref{eq:bch-gap}, the span of a weight-$b_\star$ vector, where $1\leq b_\star<n/2$ for $t\geq2$, has ordered substitution probability $R_{n,l_t}(b_\star)$ by Theorem~\ref{thm:ordered}.
This gives the first inequality since every code has $\PS\geq2^{-l_t}$.
The second follows from Corollary~\ref{cor:ord-ind-comparison} for this span, and the third from $1-b_\star/n=1/2+\lfloor(t-1)\sqrt n\rfloor/n\leq(1+\bar\rho)/2$, since $\lfloor x\rfloor\leq x$.
\end{proof}

By \eqref{eq:bch-gap}, both bounds in \eqref{eq:bch-bounds} exceed the optimum $2^{-l_t}$, and hence the exact values, by at most the factor $(1+\bar\rho)^{l_t}$, i.e., by at most $l_t\log_2(1+\bar\rho)$ bits.
For instance, at $n=2^{16}$ and $t=4$, each extra tag bit lowers the last bound in \eqref{eq:bch-gap} by only $1-\log_2(1+\bar\rho)\approx0.97$ bits.
Hence, for every subcode with $l_s\leq mt=64$ \cite[Ch.~9, Cor.~8]{macwilliams1977theory}, \eqref{eq:rho-taglength} with $\bar\rho$ in place of $\rho$ gives $l_t=67$ for $2^{-64}$.
In comparison, the first-order RM subcode achieves $2^{-64}$ at $l_t=64$ with the independent selector, but needs $n=2^{64}$ for $l_s=64$ by Proposition~\ref{prop:balanced}.
Thus, at the cost of three tag bits, the BCH-dual subcode shortens the blocklength from $2^{64}$ to $2^{16}$ and the selection bits $l_t\log_2 n$ from $4096$ to $1072$.
Next, we evaluate these subcodes exactly.

\section{Exact Evaluation and Numerical Results}
\label{sec:eval}

In this section, we evaluate the effects of coordinate order (Section~\ref{ssec:orderings}), tag length (Section~\ref{ssec:long-tags}), and balance and source dimension (Sections~\ref{ssec:tradeoff} and~\ref{ssec:code-ensemble}) on the substitution probability.
The ordered and independent probabilities follow from their exact formulas in \eqref{eq:ordered-imbalance-exact} and \eqref{eq:iid}.
For the subset selector, we use Corollary~\ref{thm:rm} for the first-order RM subcode in the canonical order.
Otherwise, we evaluate \eqref{eq:Pt} for every nonzero codeword, computing the counts in \eqref{eq:N} in exact integers by a dynamic program over the coordinates.
Displayed values are rounded to five significant digits, and Appendix~\ref{app:repro} specifies the generator matrices.

\subsection{Coordinate Order and the Weight-Set Bound}
\label{ssec:orderings}

We first compare, at $(n,l_s,l_t)=(128,7,16)$, the canonical order with the cyclic order given by the first seven rows of the BCH-dual generator.
By Section~\ref{ssec:nearbalanced}, their codes differ only by a coordinate permutation and share the nonzero weight set $\{64\}$.
Their common values are $B=H_{128,16}(64)\approx0.20988$, $\PSord\approx1.6308\times10^{-5}$, and $\PSind=2^{-16}\approx1.5259\times10^{-5}$.
As shown in Table~\ref{tab:orderings}, the canonical order meets $B$, while the cyclic order gives a value of $\PSsub$ approximately $1445$ times smaller.
In other words, the same weight-set bound is tight for one coordinate order and exceeds $\PSsub$ by more than three orders of magnitude for another.
However, the value of $\PSsub$ for the cyclic order still exceeds $\PSord$.

To distinguish a good coordinate order from the best possible one, we also enumerate all $8!=40320$ coordinate permutations of the $(8,3)$ first-order RM subcode.
For each permutation, the same order is used for all seven nonzero codewords before taking their maximum pattern probability.
The resulting extrema are given in Table~\ref{tab:optimal-ordering}, where the minimizing permutation may depend on $l_t$.
Every maximum equals $B$, as predicted by Corollary~\ref{thm:minimax}, but every minimum remains strictly above $\PSord$.
For instance, at $l_t=4$, even the best coordinate order has probability $9/35$, three times the ordered value $3/35$, and at $l_t=6$, the factor is ten.
Therefore, for this subcode, optimizing the coordinate order does not reach the substitution probability of the ordered selector, i.e., the lower bound in Theorem~\ref{thm:sandwich}.

\begin{table}[t]
\centering
\caption{Exhaustive optimization over all coordinate orders of the $(8,3)$ first-order RM subcode.
Extrema are taken separately for each $l_t$.}
\label{tab:optimal-ordering}
\footnotesize
\setlength{\tabcolsep}{7pt}
\begin{tabular}{cccc}
\toprule
$l_t$ & $\PSord$ & $\min_{\pi}\PSsub(\pi\Csub)$ & $\max_{\pi}\PSsub(\pi\Csub)=B$\\
\midrule
$2$ & $2/7$ & $3/7$ & $4/7$\\
$3$ & $1/7$ & $9/28$ & $3/7$\\
$4$ & $3/35$ & $9/35$ & $18/35$\\
$5$ & $3/70$ & $3/14$ & $3/7$\\
$6$ & $1/35$ & $2/7$ & $4/7$\\
\bottomrule
\end{tabular}
\end{table}
\begin{table}[t]
\centering
\caption{The selector with the smaller substitution probability depends on $l_t$ for the same unbalanced subcode: $n=128$, $l_s=1$, and $\rho=1/64$.}
\label{tab:long-tags}
\footnotesize
\setlength{\tabcolsep}{5pt}
\begin{tabular}{cccc}
\toprule
$l_t$ & $\PSord$ & $\PSind$ & Smaller $\PS$\\
\midrule
$125$ & $1.2721\times10^{-37}$ & $1.6328\times10^{-37}$ & Ordered\\
$126$ & $8.4808\times10^{-38}$ & $8.2914\times10^{-38}$ & Independent\\
\bottomrule
\end{tabular}
\end{table}

For other subcodes, however, a fixed coordinate order can meet this lower bound.
For instance, for the random subcode with $(n,l_s)=(128,7)$ and the generator matrix $[\mathbf I_{14}\mid\mathbf M]$ described in Appendix~\ref{app:repro}, the maximum imbalance is $\rho=1/4$.
At $l_t=8$, exact enumeration gives
\begin{equation}
\PSsub=\PSord
=\frac{\binom{80}{8}}{\binom{128}{8}}
\approx0.020275,
\label{eq:random-tight}
\end{equation}
while $\PSind=(5/8)^8\approx0.023283$.
In other words, the subset selector meets the lower bound in Theorem~\ref{thm:sandwich}.
Also, the ordered selector has a smaller substitution probability than the independent selector, as predicted by Corollary~\ref{cor:ord-ind-comparison}.
For instance, for the target $\PS\leq2^{-5}$, all three selectors first meet the target at $l_t=8$.
Indeed, consider a codeword with 80 majority bits and a tag of seven majority bits at $l_t=7$.
This tag has probability $\binom{80}{7}/\binom{128}{7}\approx0.033607>2^{-5}$ under the subset and ordered selectors, and $(5/8)^7\approx0.037253>2^{-5}$ under the independent selector.
These lower bounds increase for shorter tags.
At $l_t=8$, the subset selector achieves the substitution probability of the ordered selector with $41$ instead of $56$ selection bits.

\begin{figure}[t]
\centering
\begingroup
\pgfplotstableread{
lt subset ordered independent
1 5.00000000000000e-01 5.00000000000000e-01 5.00000000000000e-01
2 5.00488758553275e-01 2.50244379276637e-01 2.50000000000000e-01
3 3.75366568914956e-01 1.25122189638319e-01 1.25000000000000e-01
4 3.75734214917811e-01 6.26223691529685e-02 6.25000000000000e-02
5 3.13111845764843e-01 3.13111845764842e-02 3.12500000000000e-02
6 3.13419119411324e-01 1.56709559705662e-02 1.56250000000000e-02
7 2.74241729484909e-01 7.83547798528311e-03 7.81250000000000e-03
8 2.74511387036025e-01 3.92159124337178e-03 3.90625000000000e-03
9 2.47060248332422e-01 1.96079562168589e-03 1.95312500000000e-03
10 2.47303657444080e-01 9.81363720016189e-04 9.76562500000000e-04
11 2.26695019323740e-01 4.90681860008095e-04 4.88281250000000e-04
12 2.26918805127613e-01 2.45583122432482e-04 2.44140625000000e-04
13 2.10710319047069e-01 1.22791561216241e-04 1.22070312500000e-04
14 2.10918736771151e-01 6.14565083832026e-05 6.10351562500000e-05
15 1.97736315722955e-01 3.07282541916013e-05 3.05175781250000e-05
16 1.97932288285614e-01 1.53793541791464e-05 1.52587890625000e-05
17 1.86936050047524e-01 7.68967708957318e-06 7.62939453125000e-06
18 1.87121686641414e-01 3.84865665654904e-06 3.81469726562500e-06
19 1.77765602309343e-01 1.92432832827452e-06 1.90734863281250e-06
20 1.77942483505671e-01 9.63121541415008e-07 9.53674316406250e-07
21 1.69854188800868e-01 4.81560770707504e-07 4.76837158203125e-07
22 1.70023534951218e-01 2.41020445558491e-07 2.38418579101562e-07
23 1.62939220994917e-01 1.20510222779246e-07 1.19209289550781e-07
24 1.63101997439467e-01 6.03153063060960e-08 5.96046447753906e-08
25 1.56828843691796e-01 3.01576531530480e-08 2.98023223876953e-08
26 1.56985829521317e-01 1.50939204970210e-08 1.49011611938477e-08
27 1.51379192752698e-01 7.54696024851051e-09 7.45058059692383e-09
28 1.51531027449542e-01 3.77726495888340e-09 3.72529029846191e-09
29 1.46479993201224e-01 1.88863247944170e-09 1.86264514923096e-09
30 1.46627209274793e-01 9.45265301268307e-10 9.31322574615479e-10
31 1.42045108984955e-01 4.72632650634154e-10 4.65661287307739e-10
32 1.42188155418979e-01 2.36554307517799e-10 2.32830643653870e-10
33 1.38006150847832e-01 1.18277153758899e-10 1.16415321826935e-10
34 1.38145410334056e-01 5.91982525372494e-11 5.82076609134674e-11
35 1.34308037824777e-01 2.95991262686247e-11 2.91038304567337e-11
36 1.34443839683042e-01 1.48145273033056e-11 1.45519152283669e-11
37 1.30905843901910e-01 7.40726365165279e-12 7.27595761418343e-12
38 1.31038473936258e-01 3.70738423902379e-12 3.63797880709171e-12
39 1.27762512087852e-01 1.85369211951189e-12 1.81898940354586e-12
40 1.27892220221951e-01 9.27787020222704e-13 9.09494701772928e-13
41 1.24847167359524e-01 4.63893510111352e-13 4.54747350886464e-13
42 1.24974173633541e-01 2.32182713097442e-13 2.27373675443232e-13
43 1.22133851505506e-01 1.16091356548721e-13 1.13686837721616e-13
44 1.22258350844452e-01 5.81048481808583e-14 5.68434188608080e-14
45 1.19600560608703e-01 2.90524240904291e-14 2.84217094304040e-14
46 1.19722726656311e-01 1.45410498511852e-14 1.42108547152020e-14
47 1.17228503184305e-01 7.27052492559258e-15 7.10542735760100e-15
48 1.17348491416837e-01 3.63898330462106e-15 3.55271367880050e-15
49 1.15001521588501e-01 1.81949165231053e-15 1.77635683940025e-15
50 1.15119471867053e-01 9.10678898797475e-16 8.88178419700125e-16
51 1.12905635869610e-01 4.55339449398737e-16 4.44089209850063e-16
52 1.13021674549846e-01 2.27903712083438e-16 2.22044604925031e-16
53 1.10928680576700e-01 1.13951856041719e-16 1.11022302462516e-16
54 1.11042922266275e-01 5.70346055986359e-17 5.55111512312578e-17
55 1.09060012940091e-01 2.85173027993179e-17 2.77555756156289e-17
56 1.09172561973053e-01 1.42733662101849e-17 1.38777878078145e-17
57 1.07290276421794e-01 7.13668310509246e-18 6.93889390390723e-18
58 1.07401228103719e-01 3.57203166790564e-18 3.46944695195361e-18
59 1.05611207635324e-01 1.78601583395282e-18 1.73472347597681e-18
60 1.05720649301267e-01 8.93933313781566e-19 8.67361737988404e-19
61 1.04015477538343e-01 4.46966656890783e-19 4.33680868994202e-19
62 1.04123489456867e-01 2.23715398360704e-19 2.16840434497101e-19
63 1.02496559934103e-01 1.11857699180352e-19 1.08420217248550e-19
64 1.02603216083879e-01 5.59870481849628e-20 5.42101086242752e-20
}\balancedplotdata
\begin{tikzpicture}
\begin{semilogyaxis}[
width=0.85\columnwidth,height=0.65\columnwidth,
font=\footnotesize,
xlabel={Tag length $l_t$},ylabel={Substitution probability $P_S$},
xmin=1,xmax=64,ymin=1e-20,ymax=1,
xtick={1,16,32,48,64},ytick={1,1e-5,1e-10,1e-15,1e-20},
grid=major,grid style={black!12},
tick align=outside,
legend style={at={(0.03,0.03)},anchor=south west,
draw=none,fill=white,font=\footnotesize},
legend cell align=left]
\addplot[blue!75!black,thick,mark=square*,mark size=1.3pt,
mark indices={0,7,15,23,31,39,47,55,63}]
table[x=lt,y=subset]{\balancedplotdata};
\addlegendentry{Subset ($\PSsub=B$)}
\addplot[red!75!black,thick,mark=o,mark size=1.6pt,
mark options={fill=white},mark indices={3,11,19,27,35,43,51,59}]
table[x=lt,y=ordered]{\balancedplotdata};
\addlegendentry{Ordered}
\addplot[black,thick,densely dashed]
table[x=lt,y=independent]{\balancedplotdata};
\addlegendentry{Independent ($2^{-l_t}$)}
\end{semilogyaxis}
\end{tikzpicture}
\endgroup
\caption{Exact substitution probabilities for the first-order RM subcode in the canonical order at $(n,l_s)=(1024,10)$.}
\label{fig:eval-optimal}
\end{figure}
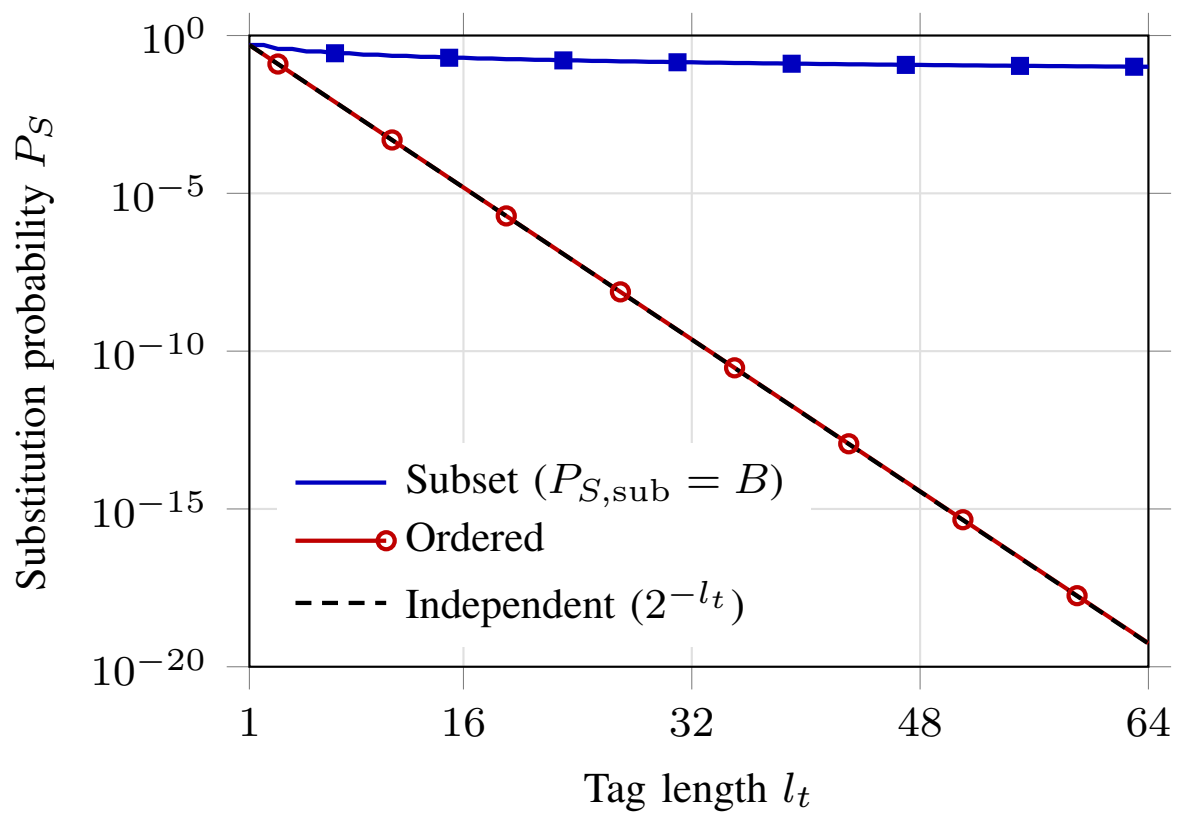

\subsection{Reversal of the Selector Comparison for Long Tags}
\label{ssec:long-tags}

Next, we show that the unbalanced case of \eqref{eq:half-block-comparison} does not extend to every $l_t<n$.
Consider the one-dimensional length-128 subcode generated by a weight-63 vector, for which $\rho=1/64$.
Table~\ref{tab:long-tags} compares the two selectors on this subcode at two tag lengths.
The ordered selector has the smaller substitution probability at $l_t=125$, but the independent selector has the smaller one at $l_t=126$.
In the latter case, $\PSord=2/\binom{128}{63}$ and $\PSind=(65/128)^{126}$.

\subsection{Exact Balance and Source Dimension}
\label{ssec:tradeoff}
In Fig.~\ref{fig:eval-optimal}, we compare the three selectors as the tag length varies for the first-order RM subcode in the canonical order at $(n,l_s)=(1024,10)$.
We observe that the independent selector achieves $2^{-l_t}$, but $\PSsub$ stays strictly above $1/(l_t+1)$ for $l_t\geq2$ despite exact balance.
For $l_t\geq2$, $\PSord$ exceeds $\PSind$, although the two curves are close over the plotted range.
For instance, at $l_t=64$, \eqref{eq:balanced-explicit} gives $\PSord\approx5.5987\times10^{-20}$, approximately $3.28\%$ above $\PSind=2^{-64}\approx5.4210\times10^{-20}$.

To illustrate the dimension limit in Proposition~\ref{prop:balanced}, we compare the first $l_s\in\{10,20,30,40\}$ rows of one length-1024 BCH-dual generator with $t=4$ in Table~\ref{tab:dimension}.
The first ten rows give a balanced subcode, while the 20-row prefix supports twice as many source bits and has $\rho=1/16$.
By \eqref{eq:rhopenalty}, its value of $\PSind$ exceeds that of the balanced prefix by the factor $(17/16)^{64}\approx48.4$, and $\PSord$ also increases.
However, once the subcode is unbalanced, the ordered selector has the smaller substitution probability, as guaranteed by Corollary~\ref{cor:ord-ind-comparison} at $l_t=64<n/2$.

For the target $\PS\leq2^{-64}$, these prefixes need at most $12$ extra tag bits with the ordered selector, while the first-order RM subcode with the same $l_s$ needs $n=2^{l_s}$ by Proposition~\ref{prop:balanced}.
Hence, for $l_s\geq20$, the selection bits drop from $64l_s$ to at most $756$, and the tag length certified by Corollary~\ref{prop:bch-guarantee} with $t=l_s/10$ is exact for $l_s\in\{20,30\}$ and exceeds the exact value by five bits for $l_s=40$.

\begin{table}[t]
\centering
\caption{Exact values for the first $l_s$ rows of a BCH-dual generator with $t=4$ at $n=1024$, and the blocklength of the first-order RM subcode with the same $l_s$, which reaches $2^{-64}$ at $l_t=64$.}
\label{tab:dimension}
\footnotesize
\setlength{\tabcolsep}{3.5pt}
\begin{tabular}{ccccccc}
\toprule
& & \multicolumn{2}{c}{$l_t=64$} & \multicolumn{2}{c}{$l_t$ for $2^{-64}$} & RM\\
\cmidrule(lr){3-4}\cmidrule(lr){5-6}\cmidrule(lr){7-7}
$l_s$ & $\rho$ & $\PSord$ & $\PSind$ & Ord. & Ind. & $n$\\
\midrule
$10$ & $0$ & $5.5987\times10^{-20}$ & $5.4210\times10^{-20}$ & $65$ & $64$ & $2^{10}$\\
$20$ & $1/16$ & $4.1342\times10^{-19}$ & $2.6250\times10^{-18}$ & $67$ & $71$ & $2^{20}$\\
$30$ & $1/8$ & $2.0041\times10^{-17}$ & $1.0182\times10^{-16}$ & $74$ & $78$ & $2^{30}$\\
$40$ & $9/64$ & $5.1024\times10^{-17}$ & $2.4616\times10^{-16}$ & $76$ & $79$ & $2^{40}$\\
\bottomrule
\end{tabular}
\end{table}
\begin{figure}[t]
\centering
\begingroup
\pgfplotstableread{
d ordlow ordmed ordhigh indlow indmed indhigh bchord bchind
4 3.49114634810711e-05 1.03811230811247e-04 3.59899331679172e-04 8.03108209098120e-05 2.15810659030693e-04 6.61303448905931e-04 1.63077810223947e-05 1.52587890625000e-05
5 4.40986907129319e-05 1.46456924471412e-04 4.47147654510487e-04 1.00452425720633e-04 2.94098011699377e-04 8.04752109493563e-04 1.63077810223947e-05 1.52587890625000e-05
6 7.20572072689831e-05 1.84252259818873e-04 5.53929183945827e-04 1.55721293069190e-04 3.61538147740297e-04 9.76989651625066e-04 1.63077810223947e-05 1.52587890625000e-05
7 1.16024316789041e-04 2.31046484534777e-04 6.84265462521316e-04 2.38593145160526e-04 4.43275722441181e-04 1.18333868549355e-03 1.63077810223947e-05 1.52587890625000e-05
8 1.84252259818873e-04 3.59899331679172e-04 1.03560673520307e-03 3.61538147740297e-04 6.61303448905931e-04 1.72431142404339e-03 4.40986907129319e-05 1.00452425720633e-04
9 2.31046484534777e-04 4.47147654510487e-04 1.26898290088263e-03 4.43275722441181e-04 8.04752109493563e-04 2.07466761382197e-03 4.40986907129319e-05 1.00452425720633e-04
10 3.59899331679172e-04 5.53929183945827e-04 1.26898290088263e-03 6.61303448905931e-04 9.76989651625066e-04 2.07466761382197e-03 4.40986907129319e-05 1.00452425720633e-04
11 4.47147654510487e-04 6.84265462521316e-04 1.26898290088263e-03 8.04752109493563e-04 1.18333868549355e-03 2.07466761382197e-03 4.40986907129319e-05 1.00452425720633e-04
12 5.53929183945827e-04 8.42935714700172e-04 1.89091972597275e-03 9.76989651625066e-04 1.43002390385314e-03 2.98456190918788e-03 4.40986907129319e-05 1.00452425720633e-04
13 6.84265462521316e-04 1.03560673520307e-03 1.89091972597275e-03 1.18333868549355e-03 1.72431142404339e-03 2.98456190918788e-03 4.40986907129319e-05 1.00452425720633e-04
14 8.42935714700172e-04 1.26898290088263e-03 2.79038424427330e-03 1.43002390385314e-03 2.07466761382197e-03 4.25895216998129e-03 4.40986907129319e-05 1.00452425720633e-04
}\vsevenA
\pgfplotstableread{
d ordlow ordmed ordhigh indlow indmed indhigh bchord bchind
4 5.75174943636217e-10 2.15383689107227e-09 2.45085958249924e-08 6.44982795520790e-09 4.70932822255181e-08 4.37322251534879e-07 2.68674701519555e-10 2.32830643653870e-10
5 7.26536770908906e-10 3.75797659110165e-09 4.01940971529875e-08 1.00906898331593e-08 8.64936404855270e-08 6.47625957734340e-07 2.68674701519555e-10 2.32830643653870e-10
6 1.27427001183439e-09 5.74749360992017e-09 6.54139228176071e-08 2.42491211151405e-08 1.30709832271485e-07 9.54508779382467e-07 2.68674701519555e-10 2.32830643653870e-10
7 2.53785432126345e-09 9.08103990367387e-09 1.05668644551519e-07 5.69266889175919e-08 1.96493366105751e-07 1.40029044458560e-06 2.68674701519555e-10 2.32830643653870e-10
8 5.74749360992017e-09 2.45085958249924e-08 2.69894406593852e-07 1.30709832271485e-07 4.37322251534879e-07 2.97324988708654e-06 7.26536770908906e-10 1.00906898331593e-08
9 9.08103990367387e-09 4.01940971529875e-08 4.26923879521185e-07 1.96493366105751e-07 6.47625957734340e-07 4.30424570784174e-06 7.26536770908906e-10 1.00906898331593e-08
10 2.45085958249924e-08 6.54139228176071e-08 4.26923879521185e-07 4.37322251534879e-07 9.54508779382467e-07 4.30424570784174e-06 7.26536770908906e-10 1.00906898331593e-08
11 4.01940971529875e-08 1.05668644551519e-07 4.26923879521185e-07 6.47625957734340e-07 1.40029044458560e-06 4.30424570784174e-06 7.26536770908906e-10 1.00906898331593e-08
12 6.54139228176071e-08 1.69468580884512e-07 1.04751498258456e-06 9.54508779382467e-07 2.04496836559139e-06 8.90760978977520e-06 7.26536770908906e-10 1.00906898331593e-08
13 1.05668644551519e-07 2.69894406593852e-07 1.04751498258456e-06 1.40029044458560e-06 2.97324988708654e-06 8.90760978977520e-06 7.26536770908906e-10 1.00906898331593e-08
14 1.69468580884512e-07 4.26923879521185e-07 2.50705660647795e-06 2.04496836559139e-06 4.30424570784174e-06 1.81386735861883e-05 7.26536770908906e-10 1.00906898331593e-08
}\vsevenB
\pgfplotstableread{
d ordlow ordmed ordhigh indlow indmed indhigh bchord bchind
4 3.75414858703210e-05 6.82542259992800e-05 1.69949586143726e-04 5.70686184566731e-05 1.00452425720633e-04 2.38593145160526e-04 1.57581836308452e-05 1.52587890625000e-05
5 4.78108987793637e-05 9.14241320844026e-05 1.69949586143726e-04 7.17243064566394e-05 1.32485787473299e-04 2.38593145160526e-04 1.57581836308452e-05 1.52587890625000e-05
6 6.82542259992800e-05 1.08405462046450e-04 2.11803244933803e-04 1.00452425720633e-04 1.55721293069190e-04 2.94098011699377e-04 1.57581836308452e-05 1.52587890625000e-05
7 8.61623115327823e-05 1.44003988816800e-04 2.63175871659782e-04 1.25258223390820e-04 2.03854117366656e-04 3.61538147740297e-04 1.57581836308452e-05 1.52587890625000e-05
8 1.08405462046450e-04 1.69949586143726e-04 2.93039800358764e-04 1.55721293069190e-04 2.38593145160526e-04 4.00456173823088e-04 1.57581836308452e-05 1.52587890625000e-05
9 1.21446720638504e-04 2.11803244933803e-04 3.26058369413273e-04 1.73436639314703e-04 2.94098011699377e-04 4.43275722441181e-04 6.82542259992800e-05 1.00452425720633e-04
10 1.69949586143726e-04 2.36183474566471e-04 3.26058369413273e-04 2.38593145160526e-04 3.26187809093982e-04 4.43275722441181e-04 6.82542259992800e-05 1.00452425720633e-04
11 1.89797712992629e-04 2.63175871659782e-04 4.02822694146934e-04 2.64986316438060e-04 3.61538147740297e-04 5.42101086242752e-04 6.82542259992800e-05 1.00452425720633e-04
12 2.36183474566471e-04 2.93039800358764e-04 4.96288214867281e-04 3.26187809093982e-04 4.00456173823088e-04 6.61303448905931e-04 6.82542259992800e-05 1.00452425720633e-04
13 2.63175871659782e-04 3.62540424732241e-04 5.50305979750795e-04 3.61538147740297e-04 4.90359444220498e-04 7.29730041220207e-04 6.82542259992800e-05 1.00452425720633e-04
14 2.93039800358764e-04 4.02822694146934e-04 6.75280238166901e-04 4.00456173823088e-04 5.42101086242752e-04 8.86959240085343e-04 6.82542259992800e-05 1.00452425720633e-04
15 3.62540424732241e-04 4.96288214867281e-04 7.47310130238037e-04 4.90359444220498e-04 6.61303448905931e-04 9.76989651625066e-04 6.82542259992800e-05 1.00452425720633e-04
16 4.02822694146934e-04 5.50305979750795e-04 7.47310130238037e-04 5.42101086242752e-04 7.29730041220207e-04 9.76989651625066e-04 6.82542259992800e-05 1.00452425720633e-04
}\vsevenC
\pgfplotstableread{
d ordlow ordmed ordhigh indlow indmed indhigh bchord bchind
4 6.38882398405996e-10 1.83412934488160e-09 1.27906533619348e-08 3.25682721255333e-09 1.00906898331593e-08 5.69266889175919e-08 2.48871705970349e-10 2.32830643653870e-10
5 9.29232135612612e-10 3.43353103979250e-09 1.27906533619348e-08 5.14437613668592e-09 1.76047215649864e-08 5.69266889175919e-08 2.48871705970349e-10 2.32830643653870e-10
6 1.83412934488160e-09 4.91576081216937e-09 2.04154839800777e-08 1.00906898331593e-08 2.42491211151405e-08 8.64936404855270e-08 2.48871705970349e-10 2.32830643653870e-10
7 3.01416847004685e-09 9.02790974483024e-09 3.23660111879281e-08 1.56896225270246e-08 4.16737022409146e-08 1.30709832271485e-07 2.48871705970349e-10 2.32830643653870e-10
8 4.91576081216937e-09 1.27906533619348e-08 4.06517100520377e-08 2.42491211151405e-08 5.69266889175919e-08 1.60365147153028e-07 2.48871705970349e-10 2.32830643653870e-10
9 6.26024240182253e-09 2.04154839800777e-08 5.09759538747775e-08 3.00802678567784e-08 8.64936404855270e-08 1.96493366105751e-07 1.83412934488160e-09 1.00906898331593e-08
10 1.27906533619348e-08 2.57268294057890e-08 5.09759538747775e-08 5.69266889175919e-08 1.06398486801532e-07 1.96493366105751e-07 1.83412934488160e-09 1.00906898331593e-08
11 1.61733054907109e-08 3.23660111879281e-08 7.97753608867088e-08 7.02177478994118e-08 1.30709832271485e-07 2.93873587705572e-07 1.83412934488160e-09 1.00906898331593e-08
12 2.57268294057890e-08 4.06517100520377e-08 1.24072806359400e-07 1.06398486801532e-07 1.60365147153028e-07 4.37322251534879e-07 1.83412934488160e-09 1.00906898331593e-08
13 3.23660111879281e-08 6.38202887093670e-08 1.54380667454826e-07 1.30709832271485e-07 2.40452384536235e-07 5.32505933059246e-07 1.83412934488160e-09 1.00906898331593e-08
14 4.06517100520377e-08 7.97753608867088e-08 2.37955164122100e-07 1.60365147153028e-07 2.93873587705572e-07 7.86696693572770e-07 1.83412934488160e-09 1.00906898331593e-08
15 6.38202887093670e-08 1.24072806359400e-07 2.94780277942303e-07 2.40452384536235e-07 4.37322251534879e-07 9.54508779382467e-07 1.83412934488160e-09 1.00906898331593e-08
16 7.97753608867088e-08 1.54380667454826e-07 2.94780277942303e-07 2.93873587705572e-07 5.32505933059246e-07 9.54508779382467e-07 1.83412934488160e-09 1.00906898331593e-08
}\vsevenD
\begin{tikzpicture}[font=\fontsize{8}{9}\selectfont]
\begin{groupplot}[
  group style={group size=2 by 2,
    horizontal sep=0.17\columnwidth,vertical sep=34pt},
  scale only axis,
  width=0.33\columnwidth,height=0.23\columnwidth,
  ymode=log,xmin=4,
  axis line style={black!60,line width=0.35pt},
  tick align=outside,tick pos=left,
  tick style={black!60,line width=0.35pt},
  major tick length=2pt,minor tick length=0pt,minor tick num=0,
  yminorgrids=false,minor y tick num=0,
  grid=major,grid style={black!10,line width=0.25pt},
  tick label style={font=\fontsize{7.5}{8.5}\selectfont},
  xticklabel style={/pgf/number format/1000 sep={}},
  title style={font=\fontsize{8}{9}\selectfont,yshift=-2pt},
  legend columns=2,legend cell align=left,
  legend style={draw=none,fill=none,
    font=\fontsize{7}{8}\selectfont,
    /tikz/every even column/.append style={column sep=8pt},
    row sep=1pt,inner sep=0pt},
  legend image post style={mark repeat=1,mark phase=1},
]
\nextgroupplot[title={(a) $n=128$, $l_t=16$},xmax=14,xtick={4,8,12,14},ymin=1e-5,ymax=1e-2,ytick={1e-5,1e-4,1e-3,1e-2},legend to name=dimensionensemblelegend]
\addplot[draw=none,name path=Aordlow,forget plot]
  table[x=d,y=ordlow]{\vsevenA};
\addplot[draw=none,name path=Aordhigh,forget plot]
  table[x=d,y=ordhigh]{\vsevenA};
\addplot[orange!80!black,fill opacity=0.12,draw=none,forget plot]
  fill between[of=Aordlow and Aordhigh];
\addplot[draw=none,name path=Aindlow,forget plot]
  table[x=d,y=indlow]{\vsevenA};
\addplot[draw=none,name path=Aindhigh,forget plot]
  table[x=d,y=indhigh]{\vsevenA};
\addplot[violet!80!black,fill opacity=0.12,draw=none,forget plot]
  fill between[of=Aindlow and Aindhigh];
\addplot[black!45,densely dotted,line width=0.5pt,forget plot]
  coordinates {(7,1e-5) (7,1e-2)};
\addplot[blue!75!black,solid,line width=0.7pt,
  mark=o,mark size=1.25pt,mark repeat=2,mark options={solid}]
  table[x=d,y=bchord]{\vsevenA};
\addlegendentry{BCH-dual, ordered}
\addplot[teal!75!black,dashed,line width=0.7pt,
  mark=triangle,mark size=1.25pt,mark repeat=2,mark options={solid}]
  table[x=d,y=bchind]{\vsevenA};
\addlegendentry{BCH-dual, independent}
\addplot[orange!80!black,solid,line width=0.7pt,
  mark=square*,mark size=1.25pt,mark repeat=2,mark options={solid}]
  table[x=d,y=ordmed]{\vsevenA};
\addlegendentry{Random, ordered}
\addplot[violet!80!black,dashed,line width=0.7pt,
  mark=diamond*,mark size=1.25pt,mark repeat=2,mark options={solid}]
  table[x=d,y=indmed]{\vsevenA};
\addlegendentry{Random, independent}
\nextgroupplot[title={(b) $n=128$, $l_t=32$},xmax=14,xtick={4,8,12,14},ymin=1e-10,ymax=1e-4,ytick={1e-10,1e-8,1e-6,1e-4}]
\addplot[draw=none,name path=Bordlow,forget plot]
  table[x=d,y=ordlow]{\vsevenB};
\addplot[draw=none,name path=Bordhigh,forget plot]
  table[x=d,y=ordhigh]{\vsevenB};
\addplot[orange!80!black,fill opacity=0.12,draw=none,forget plot]
  fill between[of=Bordlow and Bordhigh];
\addplot[draw=none,name path=Bindlow,forget plot]
  table[x=d,y=indlow]{\vsevenB};
\addplot[draw=none,name path=Bindhigh,forget plot]
  table[x=d,y=indhigh]{\vsevenB};
\addplot[violet!80!black,fill opacity=0.12,draw=none,forget plot]
  fill between[of=Bindlow and Bindhigh];
\addplot[black!45,densely dotted,line width=0.5pt,forget plot]
  coordinates {(7,1e-10) (7,1e-4)};
\addplot[blue!75!black,solid,line width=0.7pt,
  mark=o,mark size=1.25pt,mark repeat=2,mark options={solid}]
  table[x=d,y=bchord]{\vsevenB};
\addplot[teal!75!black,dashed,line width=0.7pt,
  mark=triangle,mark size=1.25pt,mark repeat=2,mark options={solid}]
  table[x=d,y=bchind]{\vsevenB};
\addplot[orange!80!black,solid,line width=0.7pt,
  mark=square*,mark size=1.25pt,mark repeat=2,mark options={solid}]
  table[x=d,y=ordmed]{\vsevenB};
\addplot[violet!80!black,dashed,line width=0.7pt,
  mark=diamond*,mark size=1.25pt,mark repeat=2,mark options={solid}]
  table[x=d,y=indmed]{\vsevenB};
\nextgroupplot[title={(c) $n=256$, $l_t=16$},xmax=16,xtick={4,8,12,16},ymin=1e-5,ymax=1e-2,ytick={1e-5,1e-4,1e-3,1e-2}]
\addplot[draw=none,name path=Cordlow,forget plot]
  table[x=d,y=ordlow]{\vsevenC};
\addplot[draw=none,name path=Cordhigh,forget plot]
  table[x=d,y=ordhigh]{\vsevenC};
\addplot[orange!80!black,fill opacity=0.12,draw=none,forget plot]
  fill between[of=Cordlow and Cordhigh];
\addplot[draw=none,name path=Cindlow,forget plot]
  table[x=d,y=indlow]{\vsevenC};
\addplot[draw=none,name path=Cindhigh,forget plot]
  table[x=d,y=indhigh]{\vsevenC};
\addplot[violet!80!black,fill opacity=0.12,draw=none,forget plot]
  fill between[of=Cindlow and Cindhigh];
\addplot[black!45,densely dotted,line width=0.5pt,forget plot]
  coordinates {(8,1e-5) (8,1e-2)};
\addplot[blue!75!black,solid,line width=0.7pt,
  mark=o,mark size=1.25pt,mark repeat=2,mark options={solid}]
  table[x=d,y=bchord]{\vsevenC};
\addplot[teal!75!black,dashed,line width=0.7pt,
  mark=triangle,mark size=1.25pt,mark repeat=2,mark options={solid}]
  table[x=d,y=bchind]{\vsevenC};
\addplot[orange!80!black,solid,line width=0.7pt,
  mark=square*,mark size=1.25pt,mark repeat=2,mark options={solid}]
  table[x=d,y=ordmed]{\vsevenC};
\addplot[violet!80!black,dashed,line width=0.7pt,
  mark=diamond*,mark size=1.25pt,mark repeat=2,mark options={solid}]
  table[x=d,y=indmed]{\vsevenC};
\nextgroupplot[title={(d) $n=256$, $l_t=32$},xmax=16,xtick={4,8,12,16},ymin=1e-10,ymax=1e-4,ytick={1e-10,1e-8,1e-6,1e-4}]
\addplot[draw=none,name path=Dordlow,forget plot]
  table[x=d,y=ordlow]{\vsevenD};
\addplot[draw=none,name path=Dordhigh,forget plot]
  table[x=d,y=ordhigh]{\vsevenD};
\addplot[orange!80!black,fill opacity=0.12,draw=none,forget plot]
  fill between[of=Dordlow and Dordhigh];
\addplot[draw=none,name path=Dindlow,forget plot]
  table[x=d,y=indlow]{\vsevenD};
\addplot[draw=none,name path=Dindhigh,forget plot]
  table[x=d,y=indhigh]{\vsevenD};
\addplot[violet!80!black,fill opacity=0.12,draw=none,forget plot]
  fill between[of=Dindlow and Dindhigh];
\addplot[black!45,densely dotted,line width=0.5pt,forget plot]
  coordinates {(8,1e-10) (8,1e-4)};
\addplot[blue!75!black,solid,line width=0.7pt,
  mark=o,mark size=1.25pt,mark repeat=2,mark options={solid}]
  table[x=d,y=bchord]{\vsevenD};
\addplot[teal!75!black,dashed,line width=0.7pt,
  mark=triangle,mark size=1.25pt,mark repeat=2,mark options={solid}]
  table[x=d,y=bchind]{\vsevenD};
\addplot[orange!80!black,solid,line width=0.7pt,
  mark=square*,mark size=1.25pt,mark repeat=2,mark options={solid}]
  table[x=d,y=ordmed]{\vsevenD};
\addplot[violet!80!black,dashed,line width=0.7pt,
  mark=diamond*,mark size=1.25pt,mark repeat=2,mark options={solid}]
  table[x=d,y=indmed]{\vsevenD};
\end{groupplot}
\path (group c1r1.north west) -- (group c1r2.south west)
  node[midway,rotate=90,anchor=south,yshift=28pt,inner sep=0pt]
  {Substitution probability $P_S$};
\path (group c1r2.south west) -- (group c2r2.south east)
  node[midway,anchor=north,yshift=-13pt,inner sep=0pt]
  {Source dimension $l_s$};
\end{tikzpicture}
\par\vspace{4pt}
\pgfplotslegendfromname{dimensionensemblelegend}
\endgroup
\caption{Exact substitution probabilities versus source dimension for BCH-dual subcodes and 100 random generators at each blocklength.
Random-subcode curves show medians, and shaded bands span the empirical 5th and 95th percentiles of the exact probabilities across random subcodes.
Each generator and its row order are fixed as $l_s$ varies, and the same generators are used for both selectors and tag lengths.
The vertical dotted line marks the exact-balance dimension limit $l_s=\log_2 n$.}
\label{fig:dimension-ensemble}
\end{figure}
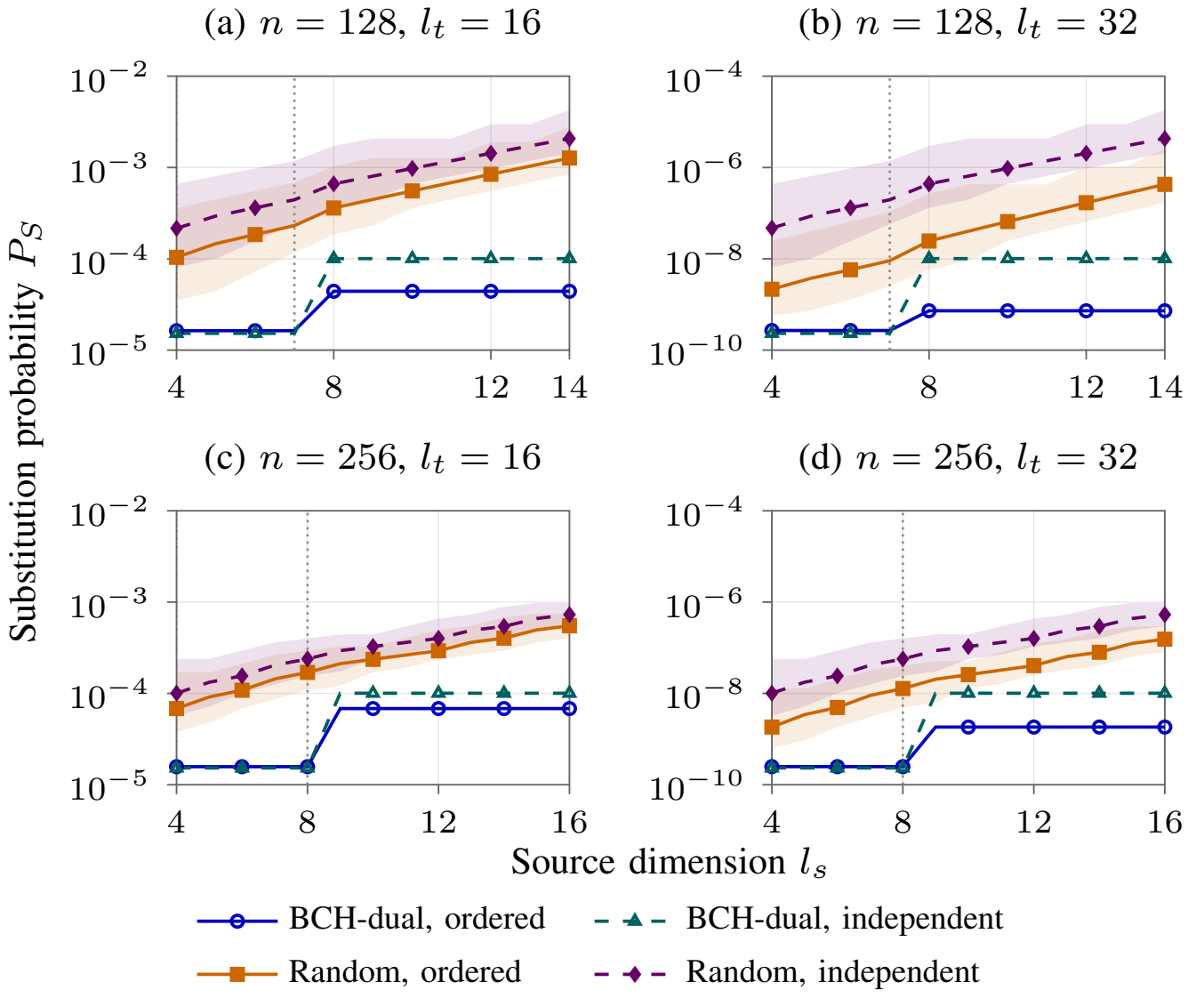

\subsection{Source Dimension Across Subcodes}
\label{ssec:code-ensemble}
For $n=2^m$ with $m\in\{7,8\}$, we compare source dimensions $4\leq l_s\leq2m$ at $l_t\in\{16,32\}$ in Fig.~\ref{fig:dimension-ensemble}.
We use one BCH-dual generator and 100 random full-row-rank $2m\times n$ generators whose spans exclude $\mathbf1_n$.
For the BCH-dual prefixes at both blocklengths, exact enumeration gives $\rho=0$ for $l_s\leq m$ and $\rho=1/8$ for $m<l_s\leq2m$.
As a result, both substitution probabilities increase at the first added row and remain constant thereafter.
At $(n,l_s,l_t)=(256,16,32)$, they are $\PSord\approx1.8341\times10^{-9}$ and $\PSind\approx1.0091\times10^{-8}$, compared with $\PSind=2^{-32}\approx2.3283\times10^{-10}$ for an exactly balanced subcode.

We also observe that the median substitution probabilities of the random subcodes increase with the source dimension.
At the largest tested dimension, the median of $\rho$ is $0.35938$ for $n=128$ and $0.27344$ for $n=256$.
At $(256,16,32)$, the medians of $\PSord$ and $\PSind$ are $1.5438\times10^{-7}$ and $5.3251\times10^{-7}$, respectively, both above the BCH-dual values.
Note that every sampled random subcode is unbalanced.
Hence, by Corollary~\ref{cor:ord-ind-comparison}, the ordered selector has the smaller substitution probability for every random subcode at both tested tag lengths.
In contrast, the independent selector has the smaller one for the balanced BCH-dual prefixes.
Increasing $l_t$ from 16 to 32 reduces both substitution probabilities, but it increases the factor $(1+\rho)^{l_t}$ in \eqref{eq:rhopenalty}.
These results are empirical and do not show which near-balanced codes are optimal.

\section{Conclusion}
\label{sec:conclusion}

In this paper, we propose systematic authentication codes that mask $l_t$ selected coordinates of a codeword from a binary linear subcode excluding the all-ones vector.
We show that all three proposed selectors achieve the optimal impersonation probability $2^{-l_t}$.
For the subset selector, we derive a bound on the substitution probability from the nonzero codeword weights, which is tight for some coordinate order.
Also, we derive exact substitution probabilities of the other two selectors in terms of the maximum weight imbalance.
We also show that, for $2\leq l_t<n$, only the independent selector can achieve a substitution probability of $2^{-l_t}$, exactly for balanced subcodes.
However, exact balance limits the source dimension to $\log_2 n$, so we also bound the ordered and independent selectors for higher-dimensional BCH-dual subcodes.
A future direction is to identify the code designs that minimize the substitution probability.
For the ordered and independent selectors, this is equivalent to minimizing the maximum weight imbalance, while the subset selector also needs the best coordinate order.
Another direction is to extend the analysis to several messages that share one coordinate selection and use fresh masks, while quantifying the security degradation under key reuse for multi-round communication.
Also, analyzing constructions with polar codes and low-density parity-check codes for 5G New Radio would identify practical use cases in communication systems.

\appendices

\section{Proof of Corollary~\ref{cor:ord-ind-comparison}}
\label{app:ord-ind-proof}

\begin{proof}
By Theorem~\ref{thm:iid}, $\PSind=((1+\rho)/2)^{l_t}$.
By the proof of Theorem~\ref{thm:ordered}, $\PSord$ is the largest probability of a projection pattern of a nonzero codeword, and a pattern with $w_t$ ones of a weight-$w$ codeword has probability $(w)_{w_t}(n-w)_{l_t-w_t}/(n)_{l_t}$.
Expanding the falling factorials, this probability is the product of $l_t$ draw probabilities, taken in any order of the pattern values.
Each of them is the number of unselected coordinates that hold the drawn value, divided by the number of unselected coordinates.
Also, $w$ and $n-w$ are at most $n(1+\rho)/2$ by \eqref{eq:rho}.

\emph{Item~1:} For $l_t=1$, both selectors pick one uniformly random coordinate.
Hence, they have the same substitution probability, which is $(1+\rho)/2$ by Theorem~\ref{thm:iid}.

\emph{Item~2:} If $\rho=0$, then $\PSind=2^{-l_t}$, and every nonzero codeword has weight $n/2$.
For such a codeword, the first two draws give a one and then a zero with probability $(n/2)^2/(n(n-1))>1/4$.
The $2^{l_t-2}$ patterns that start with these two values have this total probability, and hence one of them has probability above $2^{-l_t}$.
Thus, $\PSord>2^{-l_t}$.

\emph{Item~3:} Let $\rho>0$ and $2\leq l_t\leq\lfloor n/2\rfloor$.
We show that every pattern has probability below $((1+\rho)/2)^{l_t}$, which gives $\PSord<\PSind$.
To this end, we order the pattern as follows.
First come $|l_t-2w_t|$ unpaired draws of its more frequent value, and then the remaining draws form pairs, each consisting of a one and a zero.
The unpaired draws remove coordinates holding the same value, and hence their probabilities decrease, starting from $w/n$ or $(n-w)/n$.
Thus, each of them is at most $(1+\rho)/2$, with strict inequality from the second one on.
Next, consider a pair drawn when $r$ coordinates are unselected.
The numbers of these coordinates holding a one and a zero sum to $r$, and hence the pair has probability at most $(r/2)^2/(r(r-1))=r/(4(r-1))$.
At most $l_t-2$ draws precede the pair, and hence $r-1\geq n-l_t+1>n/2$.
Also, $n\rho$ is a positive integer by \eqref{eq:rho}, i.e., $n\rho\geq1$.
Therefore, $1/(r-1)<2/n\leq2\rho$, and the pair has probability below $(1+2\rho)/4<((1+\rho)/2)^2$.
Since $l_t\geq2$, the pattern contains a pair or at least two unpaired draws, and hence its probability is below $((1+\rho)/2)^{l_t}$.
\end{proof}

\section{Generator Matrices}
\label{app:repro}

The generator of the first-order RM subcode evaluates $x_1,\ldots,x_m$ at the $m$-bit representations of the integers $2^m-1,\ldots,0$, where $x_1$ is the most significant bit.
For the BCH-dual generators, we construct $\mathbb F_{2^m}$ using $y^7+y+1$, $y^8+y^4+y^3+y^2+1$, and $y^{10}+y^3+1$ for $m=7,8,10$, respectively, and set $\beta=y$.
Row $(d,j)$ evaluates the absolute trace $\mathrm{Tr}(\beta^j x^d)$ at $x=\beta^0,\ldots,\beta^{2^m-2}$, where the rows are ordered first by $d=1,3,\ldots,2t-1$ and then by $j=0,\ldots,m-1$.
These $tm$ rows form a basis of the dual of the primitive narrow-sense binary BCH code of designed distance $2t+1$ \cite[Ch.~9, Cor.~8 and proof of Thm.~18]{macwilliams1977theory}.
We use $t=2$, except $t=4$ at $m=10$ in Table~\ref{tab:dimension}.
Appending the identically zero coordinate, which is included in sampling, gives the length $n=2^m$.
The first $m$ rows contain each binary $m$-column once and hence generate a coordinate permutation of the first-order RM subcode.
All columns are retained in every row prefix.
For Table~\ref{tab:dimension}, we compute $\rho$ by a fast Walsh--Hadamard transform over each coset of this RM subcode, where the weight-preserving substitution $x\mapsto cx$ sets the coefficient of the highest power present, $x^5$ or $x^7$, to one.

The random example in \eqref{eq:random-tight} uses the first seven rows of $[\mathbf I_{14}\mid\mathbf M]\in\Ftwo^{14\times128}$, where $\mathbf M\in\Ftwo^{14\times114}$ has independent uniform entries.
For Fig.~\ref{fig:dimension-ensemble}, we draw 100 binary $2m\times n$ matrices with independent uniform entries at each blocklength.
We redraw any matrix that is rank-deficient or whose span contains $\mathbf1_n$, and use the first $l_s$ rows of each accepted matrix.

\begingroup
\def\IEEEbibitemsep{0pt plus 0.3pt}
\bibliographystyle{IEEEtran}
\bibliography{ref}
\endgroup

\end{document}